\documentclass[fleqn,usenatbib]{mnras}

\usepackage{newtxtext,newtxmath}

\usepackage[T1]{fontenc}
\usepackage{ae,aecompl}

\usepackage{graphicx}	
\usepackage{amsmath}	
\usepackage{amssymb}	
\usepackage{comment}
\usepackage[dvipsnames]{xcolor}
\usepackage{url}
\usepackage{subcaption}
\usepackage{natbib}

\newcommand{\brg}{Br$\upgamma$}
\newcommand{\hb}{H$\upbeta$}
\newcommand{\ha}{H$\upalpha$}
\newcommand{\feiilong}{[Fe\,\textsc{ii}]$_{1.644\,\rm \upmu m}$}
\newcommand{\feii}{[Fe\,\textsc{ii}]}
\newcommand{\hh}{H\textsubscript{2}}

\title[Jet-ISM interactions in UGC\,05771]{Searching for signs of jet-driven negative feedback in the nearby radio galaxy UGC\,05771}

\author[H. R. M. Zovaro et al.]{Henry R. M. Zovaro,$^{1}$\thanks{E-mail: henry.zovaro@anu.edu.au}
	Nicole P. H. Nesvadba$^{2}$,
	Robert Sharp$^{1}$,
	Geoffrey V. Bicknell$^{1}$,
	\newauthor
	Brent Groves$^{1}$,
	Dipanjan Mukherjee$^{3}$,
	Alexander Y. Wagner$^{4}$
	\\
	$^{1}$Research School of Astronomy and Astrophysics, The Australian National University, Canberra, ACT 2611, Australia\\
	$^{2}$Institut d'Astrophysique Spatiale, UMR 8617, Universit\'e Paris-Sud, B\^{a}t. 121, 91405 Orsay, France \\
	$^{3}$Dipartimento di Fisica Generale, Universita degli Studi di Torino, Via Pietro Giuria 1, 10125 Torino, Italy\\	
	$^{4}$University of Tsukuba, Center for Computational Sciences, Tennodai 1-1-1, 305-0006, Tsukuba, Ibaraki, Japan\\
}
\date{Accepted 2019 August 30. Received 2019 August 29; in original form 2019 April 26}

\pubyear{2019}

\begin{document}
\label{firstpage}
\pagerange{\pageref{firstpage}--\pageref{lastpage}}
\maketitle

\begin{abstract}
	Hydrodynamical simulations predict that the jets of young radio sources can inhibit star formation in their host galaxies by injecting heat and turbulence into the interstellar medium (ISM).
	To investigate jet-ISM interactions in a galaxy with a young radio source, we have carried out a multi-wavelength study of the $z = 0.025$ Compact Steep Spectrum radio source hosted by the early-type galaxy UGC\,05771. 
	Using Keck/OSIRIS observations, we detected H\textsubscript{2}\,1--0\,S(1) and [Fe\,\textsc{ii}] emission at radii of 100s of pc, which traces shocked molecular and ionised gas being accelerated outwards by the jets to low velocities, creating a `stalling wind'.
	At kpc radii, we detected shocked ionised gas using observations from the CALIFA survey, covering an area much larger than the pc-scale radio source.
	We found that existing interferometric radio observations fail to recover a large fraction of the source's total flux, indicating the likely existence of jet plasma on kpc scales, which is consistent with the extent of shocked gas in the host galaxy. 
	To investigate the star formation efficiency in UGC\,05771, we obtained IRAM CO observations to analyse the molecular gas properties. We found that UGC\,05771 sits below the Kennicutt-Schmidt relation, although we were unable to definitively conclude if direct interactions from the jets are inhibiting star formation. This result shows that jets may be important in regulating star formation in the host galaxies of compact radio sources. 
\end{abstract}




\section{Introduction}

Feedback processes driven by active galactic nuclei (AGN) are now known to be pivotal in shaping the properties of galaxies in the modern universe. For example, powerful quasar winds have been invoked to explain the observed galaxy luminosity function~\citep{Croton2006} and to establish correlations between properties of the stellar bulge and the mass of the central black hole~\citep{Silk&Rees1998}. 
Powerful, relativistic jets from radio galaxies in clusters have long been thought to prevent catastrophic cooling of the intracluster medium, reproducing observed star formation rates (SFRs)~\citep{Fabian2012,McNamaraNulsen2012}. 
In recent decades, numerous observational and theoretical studies have also shown that jets can both enhance (`positive feedback') and inhibit (`negative feedback') star formation in their host galaxies, carrying implications for the importance of radio activity in the context of galaxy evolution.
 
Hydrodynamical simulations suggest that jets have the potential to regulate star formation in the host galaxy, either by removing gas from the host galaxy altogether, or by heating it and thus preventing it from forming stars. 
When nascent jets emerge from the nucleus and propagate into the inhomogeneous interstellar medium (ISM) of their host galaxy, the jets become deflected and split as they encounter dense clumps, injecting their energy isotropically and forming a bubble that drives a shock into the ISM~\citep{Sutherland&Bicknell2007,Wagner2016}. 
While powerful jets may be able to expel gas from the host galaxy entirely, less powerful jets can create a `galactic fountain', in which gas falls back towards the nucleus after being accelerated outward~\citep{Mukherjee2016}, inducing turbulence into the ISM. 
Jets that are unable to drill efficiently through the ISM may also become temporarily trapped, injecting energy into the ISM over long periods of time. 

Jet-induced negative feedback has now been observed in a number of radio galaxies. 
Many radio galaxies exhibit powerful outflows in the ionised, atomic and molecular phases~\citep[e.g.,][]{Nesvadba2006,Nesvadba2010,Morganti2013,Tadhunter2014}. 
Warm molecular gas, traced by ro-vibrational lines of \hh{}, is also found in about 20--30 per cent of all massive radio galaxies, where jet-induced shocks are the most likely heating mechanism~\citep{Ogle2010,Willett2010,O'Dea2016}. The fact that ro-vibrational \hh{} emission is so common in these galaxies, which are generally neither gas-rich nor very actively star forming~\citep[e.g.,][]{OcanaFlaquer2010}, has previously been interpreted as evidence that jets are able to continuously inject energy into the ISM over long periods of time~\citep[e.g.,][]{Nesvadba2010}.
Although some of these \hh{}-luminous radio galaxies exhibit moderate rates of star formation, others have very low SFRs despite having large reservoirs of molecular gas~\citep{Nesvadba2010,Nesvadba2011,Guillard2015}, indicating that the jets are inducing negative feedback in their host galaxies.

Gigahertz Peak Spectrum (GPS) and Compact Steep Spectrum (CSS) sources are key in studying jet-induced feedback taking place within the host galaxy. 
They are compact ($\lesssim 20\,\rm\,kpc$) radio galaxies characterised by a convex radio spectrum, peaking at GHz frequencies in GPS sources and at frequencies of a few 100 MHz in CSS sources, with both exhibiting a steep power-law spectrum to either side of the peak~\citep{O'Dea1998}. 
A larger fraction of GPS and CSS sources are asymmetric than more extended radio sources~\citep{Orienti2016}, suggesting that the jets are interacting strongly with an inhomogeneous ISM. 
They are also believed to represent AGN with young jets; age estimates based on the expansion rate of the radio lobes~\citep[e.g.,][]{deVries2009,Giroletti2003} and based on the curvature of the radio spectrum at high frequencies~\citep{Murgia2003} are typically within the range $10^2 - 10^5\,\rm yr$.
Simulations show that jets couple most strongly to the ISM during the earliest phases of jet evolution, when the jet plasma is still confined to the host galaxy's ISM~\citep{Sutherland&Bicknell2007,Mukherjee2016,Wagner2016}. 
Observations of GPS and CSS sources have confirmed the ability of compact jets to kinematically disrupt the ISM, leaving signatures such as line ratios consistent with shocks, broad line widths and outflows of a few $1000 \,\rm km\,s^{-1}$ in some sources~\citep{Holt2008}.
Therefore, GPS and CSS sources are the ideal targets for observing jet-driven feedback processes which may have a significant impact upon the host galaxy's evolution. 

A recent study by \citet{Zovaro2019} of the nearby CSS source 4C~31.04 demonstrated that compact jets have the potential to effect feedback processes far beyond the apparent extent of the radio source. 
The authors found evidence of low-surface brightness jet plasma driving shocks and turbulence into the gas of the host galaxy on\,kpc scales, while the apparent extent of the radio jets is $< 100\,\rm\,pc$. 

To establish whether the phenomenon of low-surface brightness radio plasma is common in compact radio sources, and whether it plays an important role in jet-driven feedback, we build upon the work of \citet{Zovaro2019} by studying the radio source associated with the galaxy UGC\,05771. 
It is a CSS source with a spectral peak at $150\,\rm MHz$, believed to be approximately $9\,\rm\,pc$ in size~\citep{deVries2009}, and therefore a very young radio source. 
To search for shock-heated ionised and molecular gas within a few 100\,pc of the nucleus, both important tracers of jet-ISM interactions, we obtained high-resolution, adaptive-optics-assisted near-IR integral field spectrograph observations using OSIRIS on the Keck I telescope. 
We complemented this data with optical integral field spectroscopy from the CALIFA survey, which has a much larger field of view, enabling us to search for signatures of jet-ISM interaction on\,kpc scales using the optical emission line gas.
To determine if the jets are inducing negative feedback in UGC\,05771, we also obtained CO data from the IRAM 30\,m telescope, allowing us to investigate the star formation efficiency of the host galaxy.
 
In Section~\ref{sec: Host galaxy properties}, we summarise the properties of UGC\,05771 and its radio source. In Section~\ref{sec: OSIRIS data} we discuss our OSIRIS observations, data reduction and analysis; in Section~\ref{sec: CALIFA data} we discuss the CALIFA data and our analysis; and in Section~\ref{sec: IRAM data} we discuss our IRAM observations, data reduction and analysis. We present our interpretation of our findings in Section~\ref{sec: Discussion} before summarising in Section~\ref{sec: Conclusion}.
For the remainder of this paper, we assume a cosmology with $H_0 = 70\rm\,km\,s^{-1}\,Mpc^{-1}$, $\Omega_M = 0.3$, and $\Omega_\Lambda = 0.7$, and use the fundamental plane-derived angular diameter distance and luminosity distance estimates of \citet{Saulder2016} (Table~\ref{tab: Properties of UGC05771}).

\section{Host galaxy and radio source properties}\label{sec: Host galaxy properties}

UGC\,05771 is an early-type galaxy (ETG) with stellar mass $M_* = 10^{11.27 \pm 0.10} \,\rm M_\odot$, and is the most massive galaxy in a three-member group, with the stellar mass of the other two galaxies $\sim 10^{10}\,\rm M_\odot$~\citep{Saulder2016}. 
We summarise its properties in Table~\ref{tab: Properties of UGC05771}.

\begin{table}
	\caption{Properties of UGC\,05771.
		$^a$luminosity-weighted stellar velocity dispersion within the effective radius. 
		$^b$using the $M_{\rm BH} - \sigma_*$ relation of \citet{Gultekin2009}. 
		$^c$from a S\'ersic profile fit to the $V$ band continuum.}
	\begin{tabular}{c c c}
		\hline
		\textbf{Property} & \textbf{Value} & \textbf{Reference} \\
		\hline
		$z$ & $0.02469 \pm 0.00017$ & \citet{Falco1999} \\
		$D_L$ & $95.6524\,\rm Mpc$ & \citet{Saulder2016} \\
		$D_A$ & $91.5876\,\rm Mpc$ & \citet{Saulder2016} \\
		Angular scale & $0.444\,\rm\,kpc\,arcsec^{-1}$ & \citet{Saulder2016} \\
		$\log [M_*/ \rm M_\odot]$ & $11.27 \pm 0.10$ & \citet{Sanchez2016} \\
		$\sigma_*$ & $226 \pm 3\,\rm km\,s^{-1}$ & \citet{Oh2011} \\
		$\sigma_e$ & $223 \pm 3\,\rm km\,s^{-1}$ & This work$^a$ \\
		$\log [M_{\rm BH} / \rm M_\odot]$ & $8.54 \pm^{0.23}_{0.03}$ & This work$^b$ \\
		$n$ & 1.25 & This work$^c$ \\
		$R_e$ & 3.0\,kpc & This work$^c$ \\
		\hline
	\end{tabular}
	\label{tab: Properties of UGC05771}
	
\end{table}

\begin{figure}
	\centering
	\includegraphics[width=1\linewidth]{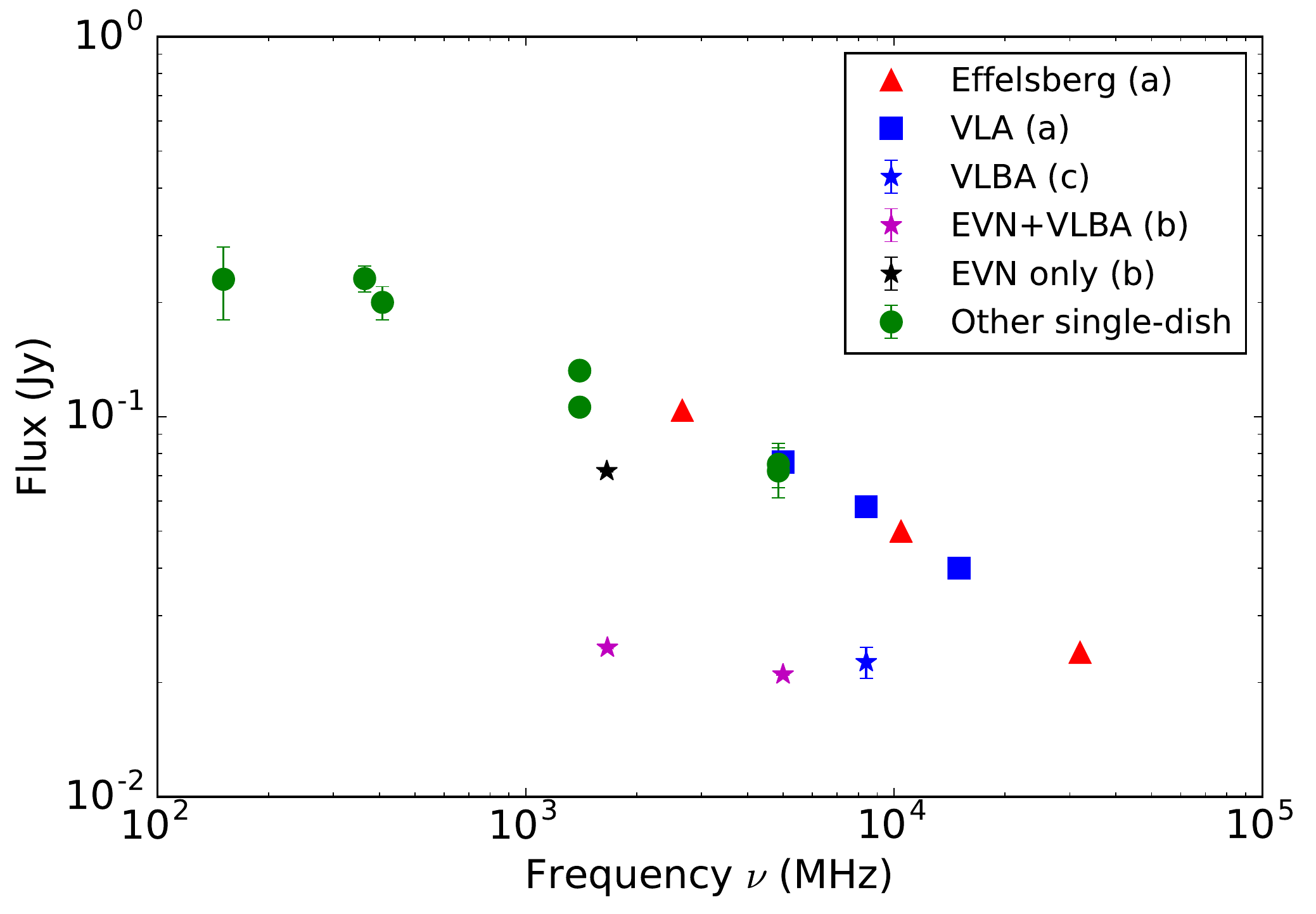}
	\caption{Radio spectrum of UGC\,05771. References: (a) \citet{Snellen2004}; (b) \citet{deVries2009}; other single dish fluxes are the integrated fluxes from the 6C 151 MHz, Texas 365 MHz, Bologna 408 MHz, Greenbank 1.4 GHz, NVSS, Becker \& White 4.85 GHz, and 87GB catalogues.}
	\label{fig: radio spectrum}
\end{figure}

UGC\,05771 hosts a low-power CSS source with $P_{5.0\,\rm GHz} = 10^{22.96}\,\rm W\,Hz^{-1}$~\citep{Snellen2004}.
Fig.~\ref{fig: radio spectrum} shows the radio spectrum obtained from the integrated fluxes from the 6C 151 MHz, Texas 365 MHz, Bologna 408 MHz, Greenbank 1.4 GHz, NVSS, Becker \& White 4.85 GHz, and 87GB catalogues. 
We also show integrated fluxes given by \citet{Snellen2004} from Jansky Very Large Array (VLA) (in BC configuration), Effelsberg observations and integrated fluxes provided by \citet{deVries2009} using European Very Long Baseline Interferometry (VLBI) Network (EVN) and Global VLBI observations, and Very Long Baseline Array (VLBA) observations provided by \citet{Cheng2018}. These interferometric observations recover only $\sim 20-30\,\rm per\,cent$ of the flux measured using single-dish observations at similar frequencies; excluding the EVN and EVN/VLBI fluxes, the spectrum is well-described by a power law at high frequencies with a spectral index $\alpha = 0.62$\footnote{We define the spectral index $\alpha$ such that $S \propto \nu^{-\alpha}$.}~\citep{Snellen2004}.
The spectrum flattens and turns over at approximately 150\,MHz.

To constrain the jet power we used the empirical correlation between radio luminosity $L_{151\rm\,MHz}$ at 151\,MHz and jet power $L_{\rm jet}$ of \citet{Ineson2017}:
\begin{equation}
L_{\rm jet} = 5 \times 10^{39} \left(\frac{L_{151\rm\,MHz}}{10^{28}\,\rm W\,Hz^{-1}\,Sr^{-1}}\right)^{0.89 \pm 0.09} \,\rm erg\,s^{-1}
\end{equation} 
Although the correlation between jet power and radio luminosity is believed to be fairly tight for extended radio sources, \citet{Ineson2017} warn that the correlation may not hold for compact radio sources, in which the radio luminosity is expected to increase over time due to the accumulation of radiative plasma.
With this caveat in mind, using the 151 MHz flux $F_{\rm 151\,MHz} = 0.230\,\rm Jy$ from the 6C survey, we estimate a jet power $L_{\rm jet} \geq 4.2\times 10^{41}\,\rm erg\,s^{-1}$. 

VLBI observations at 1.665\,GHz (EVN and VLBA) and 4.993\,GHz (EVN only), taken in 2004, reveal a symmetric core-jet structure 9\,pc across with a position angle of approximately 120$^\circ$ (\citet{deVries2009}, Figs.~\ref{fig: VLBI: 1.665 GHz} and \ref{fig: VLBI: 4.993 GHz}).
Interestingly, VLBA observations taken in 2011 at 8.4\,GHz (Fig.~\ref{fig: VLBI: 8.4 GHz}) reveal a single-sided core-jet morphology, with a bright hotspot $18.3\,\rm\,pc$ from the core to the South-East~\citep{Cheng2018}, and show no hints of emission along the jet axis. 
Although these inconsistencies are curious, they are only relevant to the parsec-scale structure of the source, which is well below the resolution limit of the observations discussed in this paper. Hence, determining the cause of these discrepancies is beyond the scope of this paper; none the less, we discuss these inconsistencies in Appendix~\ref{appendix: The morphology of the radio source}.

\section{OSIRIS observations}\label{sec: OSIRIS data}

\subsection{Observations}\label{subsec: OSIRIS observations}
We observed UGC\,05771 using the OH-Suppressing Infra-Red Imaging Spectrograph (OSIRIS)~\citep{Larkin2006} on the 10\,m Keck I telescope on Mauna Kea in Hawai`i.
OSIRIS is a lenslet-array integral field spectrograph operating over the wavelength range $1-2.4\,\rm \upmu m$.
The instrument has a configurable plate scale (0.020'', 0.035'', 0.050'' or 0.100'' per spaxel) and a wide selection of both broad- and narrow-band filters, which determine the field-of-view.
OSIRIS is fed by the Keck Adaptive Optics (AO) System~\citep{Wizinowich2006,vanDam2006}, which can be used in laser guide star (LGS) or natural guide star (NGS) mode, to provide near-diffraction limited angular resolution.

We used OSIRIS and the Keck AO system in laser guide star (LGS) mode with a position angle of 0$^\circ$ on February 12, 2018 during program Z260. We observed with the $H$ and $K$ narrow-band Hn4 ($1652-1737\,\rm nm$), Kn3 ($2121-2229\,\rm nm$) and Kn4 ($2208-2320\,\rm nm$) filters in the 0.05'' setting, providing fields of view of approximately 2.1'' $\times$ 3.2'', 2.4'' $\times$ 3.2'' and 2.1'' $\times$ 3.2'' respectively. 
The median spectral resolutions in the $K$- and $H$-bands are approximately 3900 and 2800 respectively.
We used 600\,s exposure times for both source and sky frames, integrating on-source for a total of 40, 60, and 40 minutes in the Hn4, Kn3 and Kn4 bands respectively. 
The HIPPARCOS stars HIP41798, HIP55182 and HIP53735, observed before and after UGC\,05771, were used as telluric and flux standards.

\subsection{Data reduction}\label{subsec: OSIRIS data reduction}
We reduced our observations using the \textsc{OSIRIS Data Reduction Pipeline} (ODRP)\footnote{Available \url{https://github.com/Keck-DataReductionPipelines/OsirisDRP}}, a software package for \textsc{idl}, as follows.

We mean-combined our dark frames to create a master dark, which we subtracted from all science and sky frames. 
To associate each spectra on the detector with its corresponding lenslet, we ran the `extract spectra' module, using the appropriate \textit{rectification matrix} for the plate scale and filter, which provides the point spread function (PSF) for each lenslet as a function of wavelength. 
We then resampled the spectra to a linear wavelength grid, and assembled them into a data cube, before removing sky lines using the scaled sky subtraction method of \citet{Davies2007}.
Next, we extracted 1D spectra of our telluric and flux standard stars and removed H\,\textsc{i} absorption features. We divided this 1D spectrum by a blackbody corresponding to the star's effective temperature, which we estimated using its spectral class, and normalised it to a value of 1 at the filter's effective wavelength. Lastly, we removed telluric absorption features from our object exposures by dividing the data cubes by the resulting 1D spectrum of the telluric standard.

Due to telescope pointing errors, the spatial shifts between individual exposures were not integer multiples of the spaxel size, meaning we could not use the in-built mosaicing function of the ODRP to combine our exposures.
We instead created the mosaics as follows. 
For each data cube, we generated a continuum image by summing the data cube along the wavelength axis. 
We then calculated the spatial offsets for each data cube by cross-correlating the continuum images. 
To shift each data cube by the required amount, we used a 3rd-order spline function to shift and interpolate each wavelength slice of the data cube.
We then median combined the shifted data cubes. The associated variance of each pixel in the final data cube was calculated as the variance of the pixels contributing to the pixel value.

To flux calibrate our science exposures, we generated a 1D spectrum of the flux standard in the same fashion as for the telluric standard, and multiplied it by the normalised blackbody to restore its original spectral shape, before removing telluric absorption lines by dividing the spectrum by the 1D spectrum of the telluric standard.
Then, we determined the median number of counts $n$ per second in the telluric-corrected spectrum of the flux standard in a small window around the effective wavelengths of the 2MASS $Ks$ band (for our Kn3 and Kn4 band observations) and $H$ band (for our Hn4\,band observations). 
We created a conversion factor $T$ by dividing the expected $F_\uplambda$ for the flux standard, estimated using its 2MASS magnitude, by $n$. Lastly, we multiplied our data cubes by $T$ to convert them from units of $\rm counts\,s^{-1}$ to $\rm erg\,s^{-1}\,cm^{-2}$\,\AA$\rm ^{-1}$. 

We used a Median Absolute Deviation (MAD) smoothing algorithm to smooth the mosaiced data cubes and to remove artefacts.
For each pixel, we computed the median and standard deviation of the surrounding 8 pixels, and rejected those pixels with absolute value greater than 3 standard deviations from the median. We iterated until no more pixels were rejected. The value of the central pixel was then replaced by the mean of the remaining pixels, and the variance of the central pixel was replaced by the mean of the variance of the remaining pixels.

MAD smoothing degrades the intrinsic spatial resolution of our data $\sigma$ by an amount $\sigma_{\rm MAD}$ such that the effective Gaussian sigma $\sigma^\prime$ of the PSF is increased to 
\begin{equation}
\sigma^\prime = \sqrt{\sigma_{\rm MAD}^2 + \sigma^2} 
\end{equation}
where we measured $\sigma$ by fitting a 2D Gaussian to our unsmoothed standard star exposures. 
We found the PSF of our observations to be slightly asymmetric, with $\sigma \approx 0.034\text{''} \times 0.057\text{''}$ in the Kn3 and Kn4 bands and $\sigma \approx 0.027\text{''} \times 0.041\text{''}$ in the Hn4 band.
We measured $\sigma_{\rm MAD} \approx 0.0459\text{''}$ by applying the MAD smoothing algorithm to 2D Gaussian profiles with known widths.

\subsection{Emission line fitting}\label{subsec: OSIRIS Emission line fitting}

We used \textsc{mpfit}~\citep{Markwardt2009}, a \textsc{python} implementation of the Levenberg-Marquardt algorithm~\citep{More1978} developed by M. Rivers\footnote{Available \url{http://cars9.uchicago.edu/software/python/mpfit.html}.}, to simultaneously fit single-component Gaussian profiles to emission lines and a linear component to fit to the continuum, leaving the slope and the intercept as free parameters.
We discard fits with $\chi^2 > 2$ and signal-to-noise $\rm (S/N) < 3$.
We have corrected our line widths for instrumental resolution by subtracting the width of the line spread function in quadrature from the width of the fitted Gaussian, where we estimated the width of the line spread function by fitting a Gaussian to sky lines close in wavelength to the relevant emission line.

To calculate integrated line fluxes, we summed the fluxes in each spaxel. 
Due to poor S/N in the \hh 1--0 S(0) line, we measured the total flux by fitting a Gaussian to the co-added spectrum within the central 200\,pc, corresponding to the area in which we detect \hh 1--0 S(1). 

\begin{table}
	\centering
	\caption{Total emission line fluxes and upper limits measured from the CALIFA data (top half of the table) and our OSIRIS observations (bottom half). 
	For emission lines in the CALIFA and OSIRIS data, we measured the flux by simultaneously fitting Gaussian profiles to all lines in the integrated spectrum extracted from spaxels within 10\,kpc and 200\,pc of the nucleus respectively. Upper limits were estimated using the method detailed in \ref{subsec: OSIRIS Emission line fitting}. Emission lines in the CALIFA data have been corrected for extinction (Section~\ref{subsubsec: CALIFA reddening}).
}
	\begin{tabular}{c c}
		\hline
		\textbf{Emission line} & \textbf{Flux} ($10^{-14} \,\rm erg\,s^{-1}\,cm^{-2}$) \\ 
		\hline 
		{[}O\,\textsc{ii}{]}$\uplambda\uplambda$3726,3729 & $7.56 \pm 0.12$ \\
		\hb{}											 & $2.18 \pm 0.05$ \\
		{[}O\,\textsc{iii}{]}$\uplambda\uplambda$4959,5007	& $4.35 \pm 0.05$ \\
		{[}N\,\textsc{ii}{]}$\uplambda\uplambda$6548,6583 	& $7.12 \pm 0.04$ \\
		\ha{}											& $5.64 \pm 0.03$ \\
		{[}S\,\textsc{ii}{]}$\uplambda$6716		& $4.10 \pm 0.03$ \\
		{[}S\,\textsc{ii}{]}$\uplambda$6731		& $2.25 \pm 0.03$ \\
		\hline
		\hh{}\,1--0\,S(1) & $1.85 \pm 0.04$ \\
		\hh{}\,1--0\,S(0) & $0.40 \pm 0.05$ \\ 
		\hh{}\,2--1\,S(1) & $\leq 1$ \\
		\brg{} & $\leq 0.3$ \\ 
		\feii{} & $1.09 \pm 0.05$ \\
		\hline
	\end{tabular}
	\label{tab: eline fluxes}
\end{table}

To calculate upper limits for the fluxes of emission lines not detected using our $\chi^2$ and S/N criteria, we used the following method.
In each spaxel, we calculated the standard deviation $\sigma$ of the continuum in a window centred on the emission line. 
We assumed the non-detected emission line in that spaxel was a Gaussian with amplitude $3\sigma$. 
For the \brg{} line, we used the width of the \ha{} line measured using the CALIFA data (Section~\ref{sec: CALIFA data}), and for non-detected \hh{} lines, we used the width of the \hh{}\,1--0\,S(1) line. 
To estimate an upper limit for the total emission line flux, we assumed the lines are detected in every spaxel in which we detect the \hh{}\,1--0\,S(1) emission line.

\subsection{Results}\label{subsec: OSIRIS results}
\begin{figure}
	\centering
	\subcaptionbox{\label{fig: Kn3 spectrum}}{\includegraphics[width=1\linewidth]{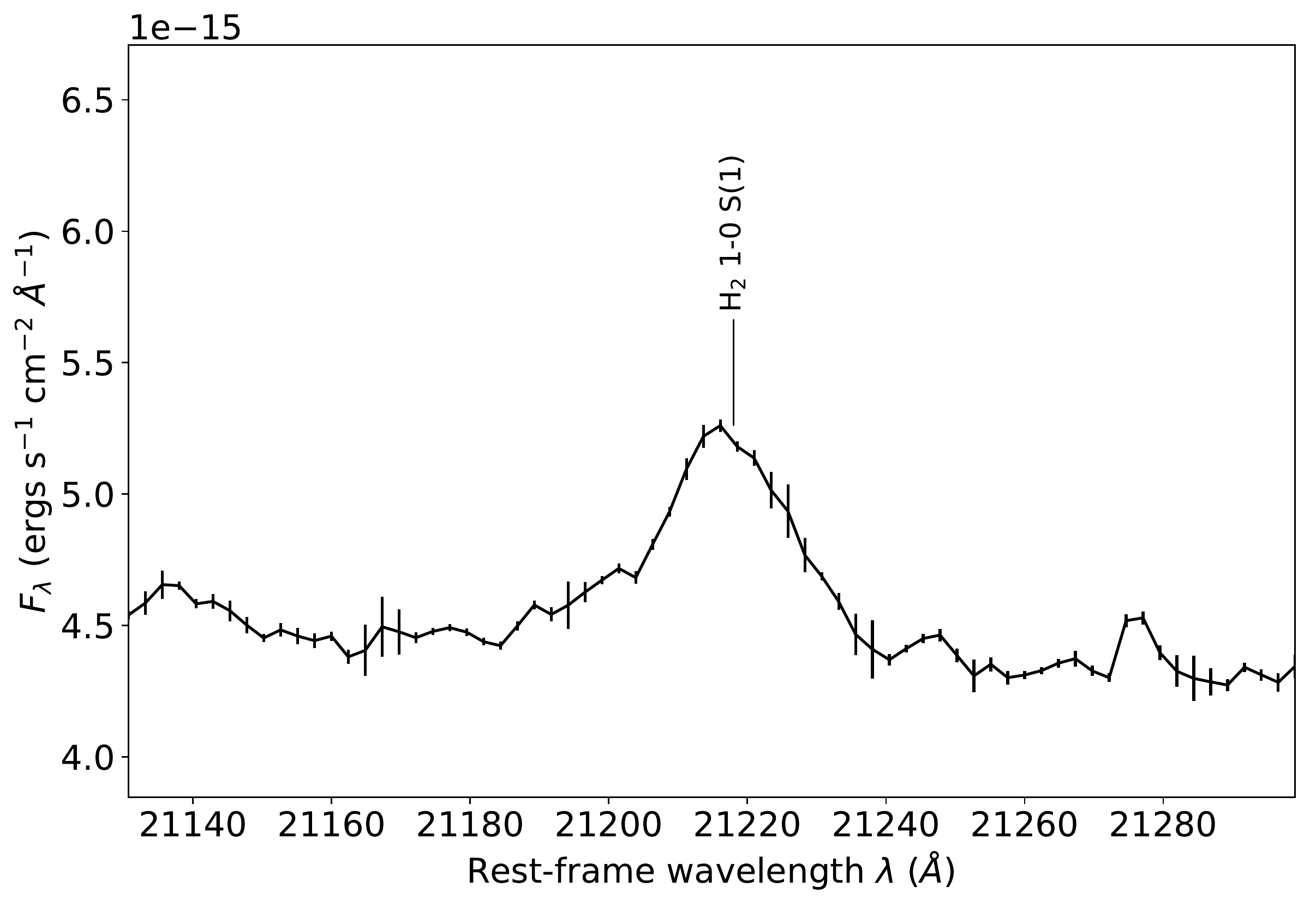}}
	\subcaptionbox{\label{fig: Hn4 spectrum}}{\includegraphics[width=1\linewidth]{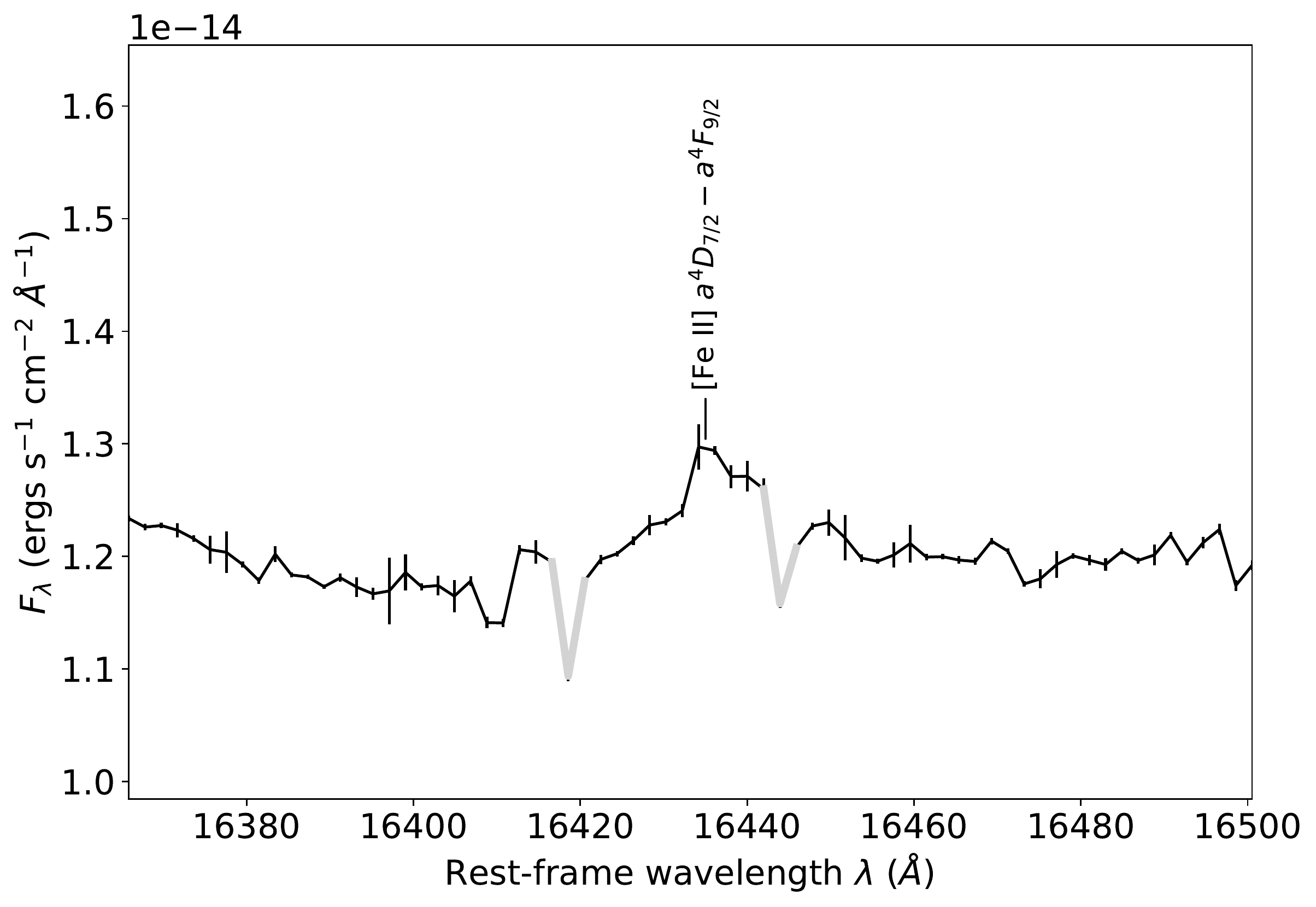}}
	\caption{Integrated spectra extracted from the OSIRIS data cubes from spaxels within 200\,pc of the nucleus in the Kn3 (a) and Hn4 (b) bands with $1\sigma$ error bars shown. Spectral regions dominated by sky emission have been indicated in grey.}
	\label{fig: OSIRIS spectra}
\end{figure}

Fig.~\ref{fig: OSIRIS spectra} shows the integrated spectra of our OSIRIS data cubes within 200\,pc of the nucleus in the Kn3 and Hn4 bands.
In the Kn3 and Kn4 bands, we detected ro-vibrational \hh{}\,1--0 emission lines which trace warm molecular gas, and in the Hn4\,band we detected the [Fe\,\textsc{ii}] $a^4D_{7/2}-a^4F_{9/2}$ line (rest-frame wavelength 1.644\,$\upmu$m) which traces the warm ionised medium.
Table~\ref{tab: eline fluxes} lists the fluxes of the detected emission lines. 

\subsubsection{\hh{} emission}
\label{sec: H2 emission}

\begin{figure*}
	\centering	
	\subcaptionbox{\label{fig: Kn3 continuum}}{\includegraphics[width=0.49\linewidth]{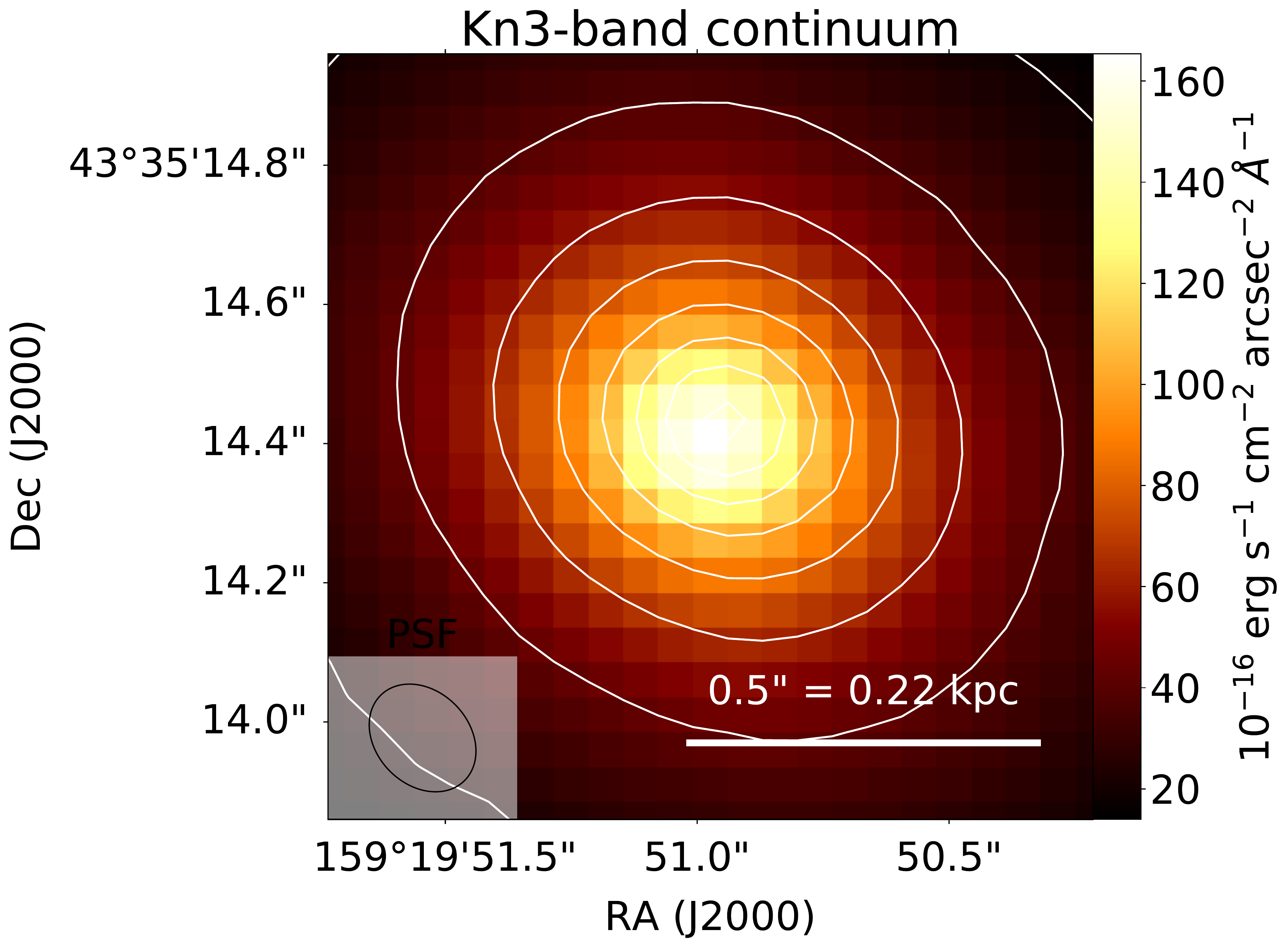}}
	\hfill
	\subcaptionbox{\label{fig: H2 flux}}{\includegraphics[width=0.49\linewidth]{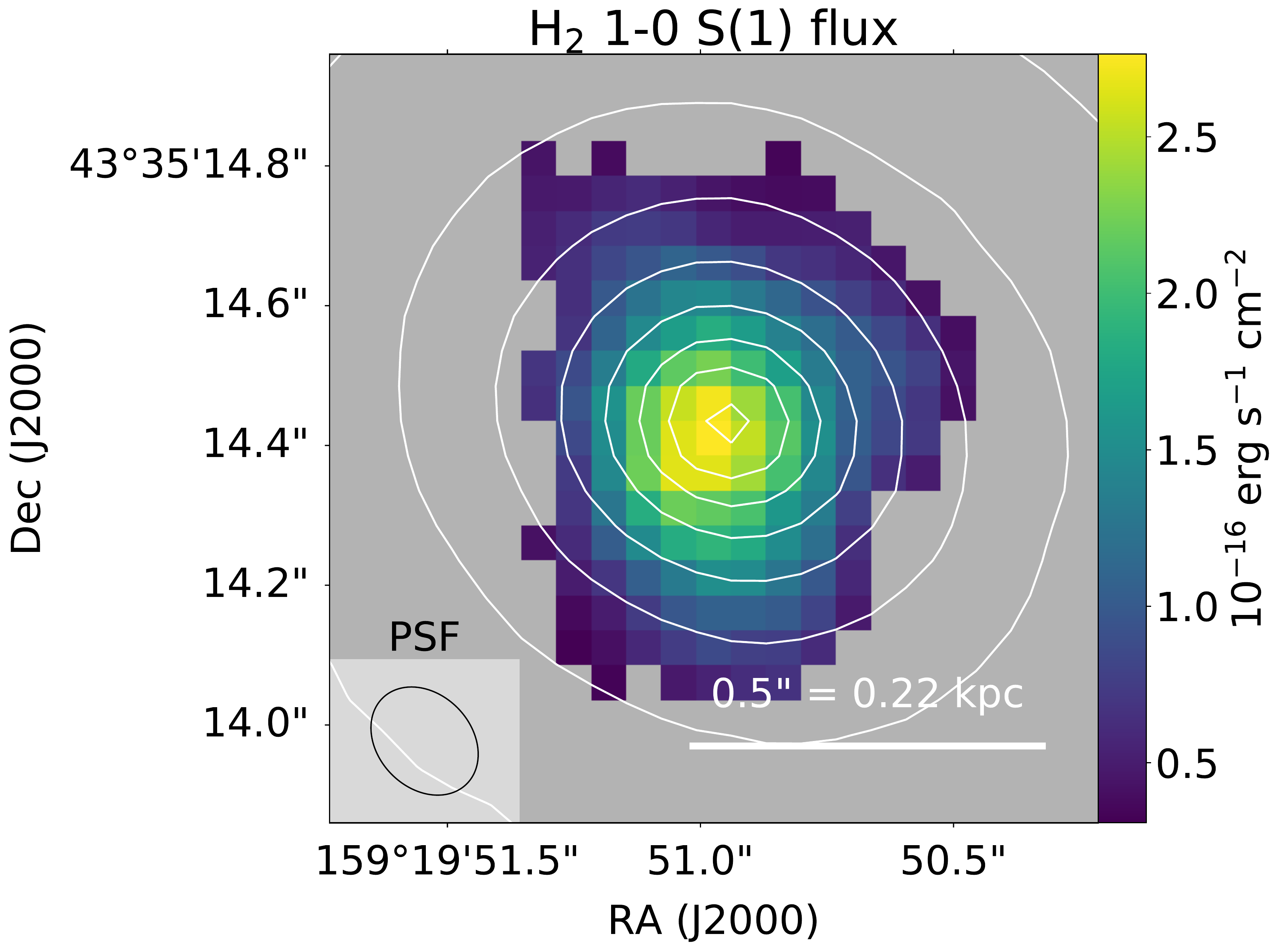}}
	\subcaptionbox{\label{fig: H2 vrad}}{\includegraphics[width=0.49\linewidth]{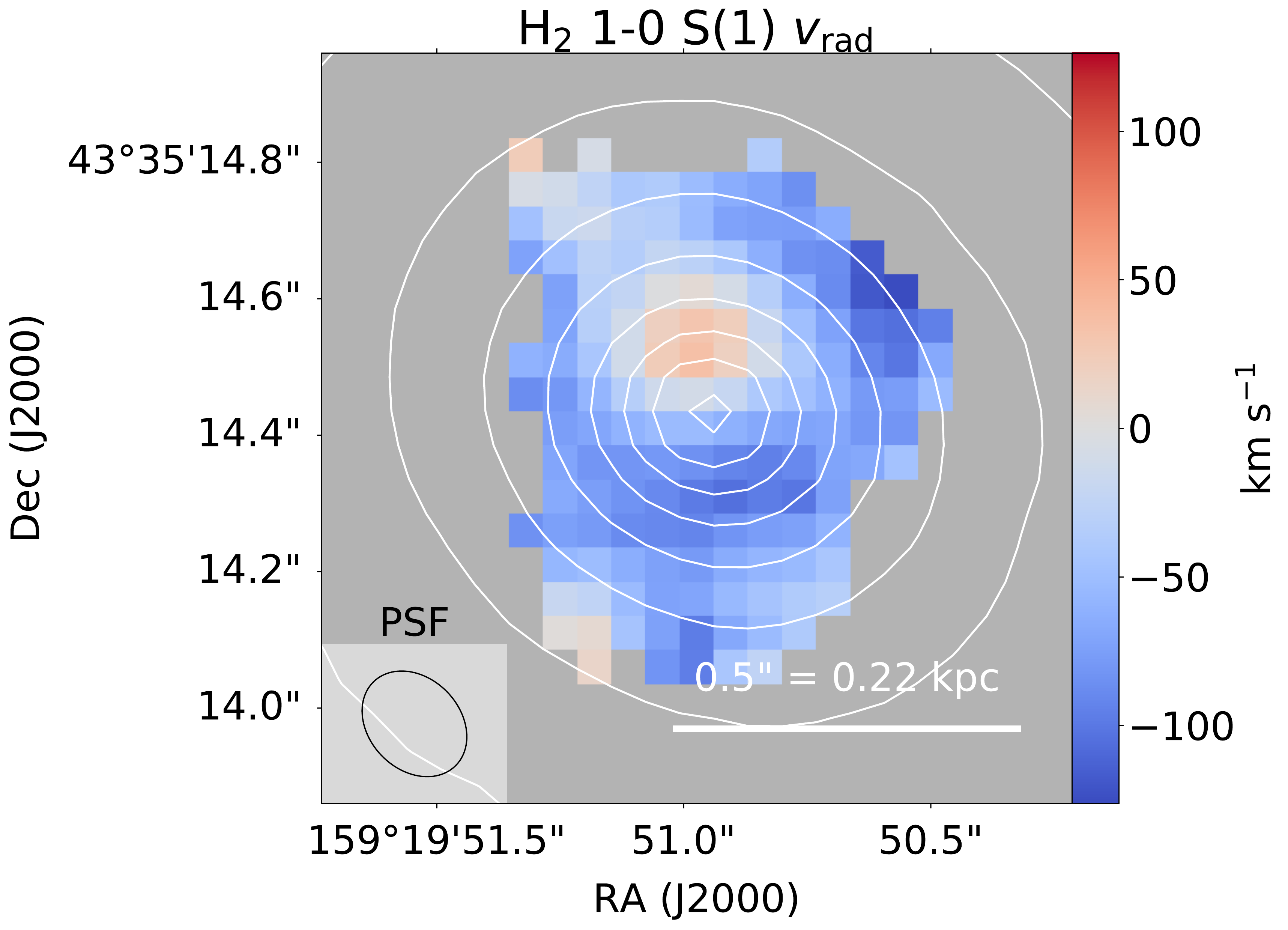}}
	\hfill
	\subcaptionbox{\label{fig: H2 vdisp}}{\includegraphics[width=0.49\linewidth]{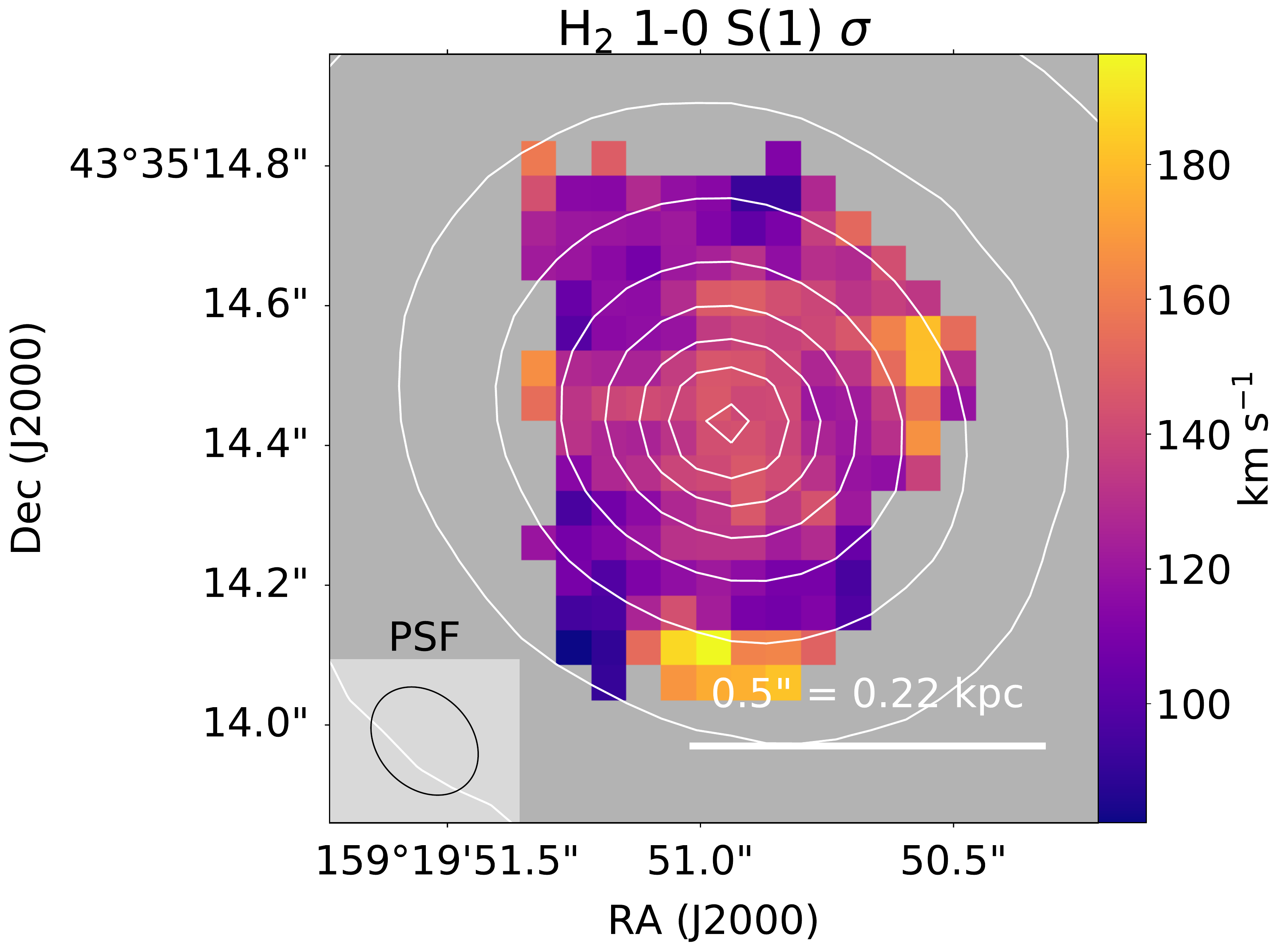}}
	\caption{(a) The Kn3\,band continuum, and (b) the integrated flux, (c) radial velocity (relative to rest frame wavelength) and (d) velocity dispersion of the \hh{}\,1--0\,S(1) emission line. The Kn3\,band continuum is indicated in contours, and the full width at half-maximum (FWHM) of the PSF (taking into account the effects of MAD smoothing) is indicated in all figures.}
	\label{fig: H2 maps}
\end{figure*}

Fig.~\ref{fig: H2 maps} shows the Kn3\,band continuum and the \hh{}\,1--0\,S(1) emission line flux (rest-frame wavelength 2.122\,$\upmu$m), radial velocity and velocity dispersion.

\begin{figure}
	\centering
	\includegraphics[width=1\linewidth]{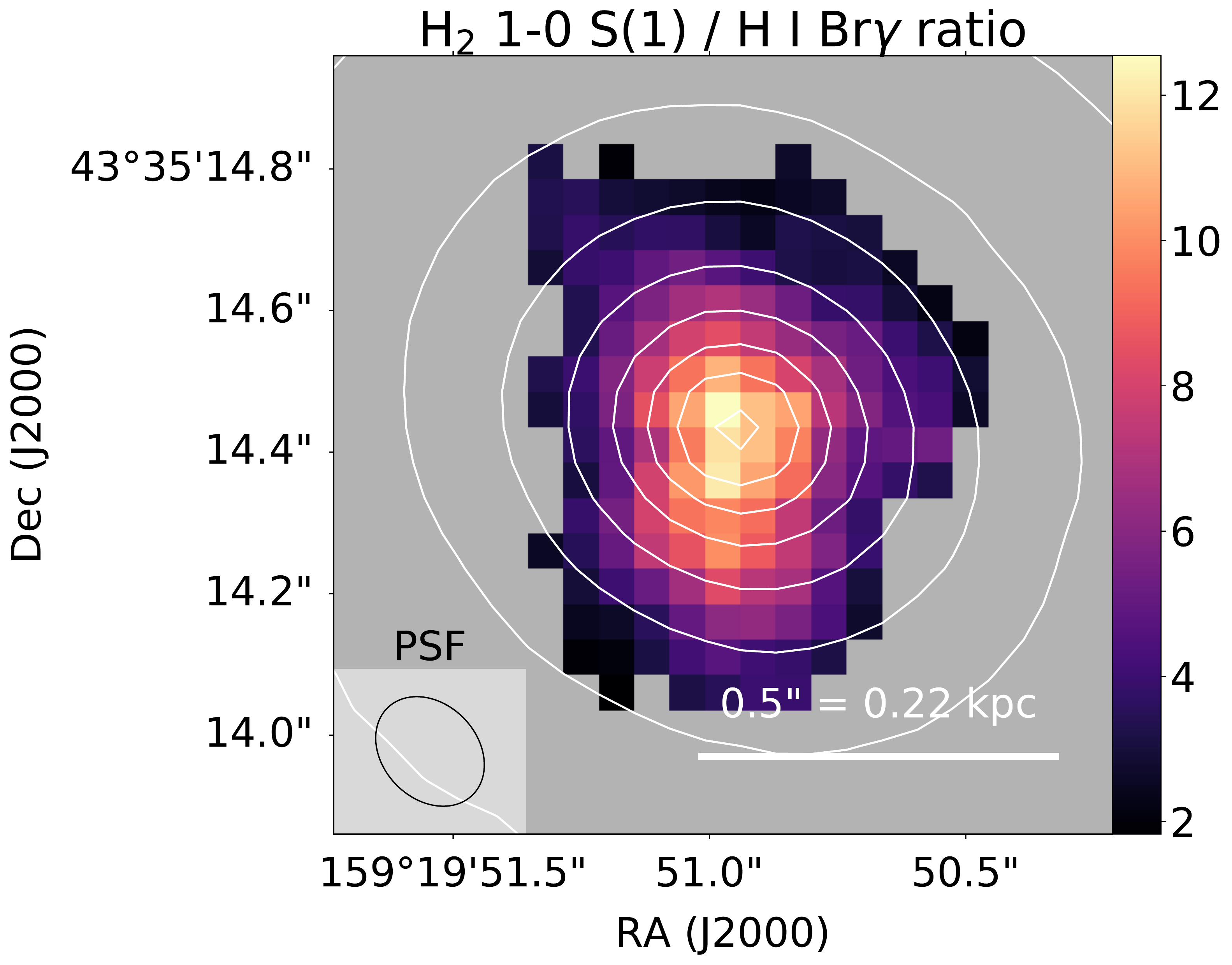}
	\caption{A map showing the \hh{}\,1--0\,S(1)/\brg{} ratio in each spaxel. The large values of this ratio suggest shocks are the excitation mechanism. The Kn3\,band continuum is indicated in contours, and the FWHM of the PSF (taking into account the effects of MAD smoothing) is indicated.}
	\label{fig: H2 1-0 S(1) to Brgamma line ratio}
\end{figure}

Ro-vibrational emission arises from several processes: hydrogen molecules can be collisionally excited by shocks, or radiatively excited by UV radiation from young stars or AGN. 
The \hh{}\,1--0\,S(1)/\brg{} ratio can be used to distinguish between shock and UV excitation.
In Fig.~\ref{fig: H2 1-0 S(1) to Brgamma line ratio} we show the \hh{}\,1--0\,S(1)/\brg{} ratio in each spaxel, where we have used upper limits for the \brg{} flux in each spaxel estimated using the method described in Section~\ref{subsec: OSIRIS Emission line fitting}. 
The ratio far exceeds the values of 0.1--1.5 typical of regions dominated by UV excitation, strongly suggesting that the \hh{} is shock heated~\citep{Puxley1990}.
The fact that \feii{} emission, a tracer of shocked gas, is present in the same region (Section~\ref{sec: [Fe II] emission}) further supports our hypothesis that the \hh{} is shock-heated. 

\begin{figure}
	\centering
	\includegraphics[width=1\linewidth]{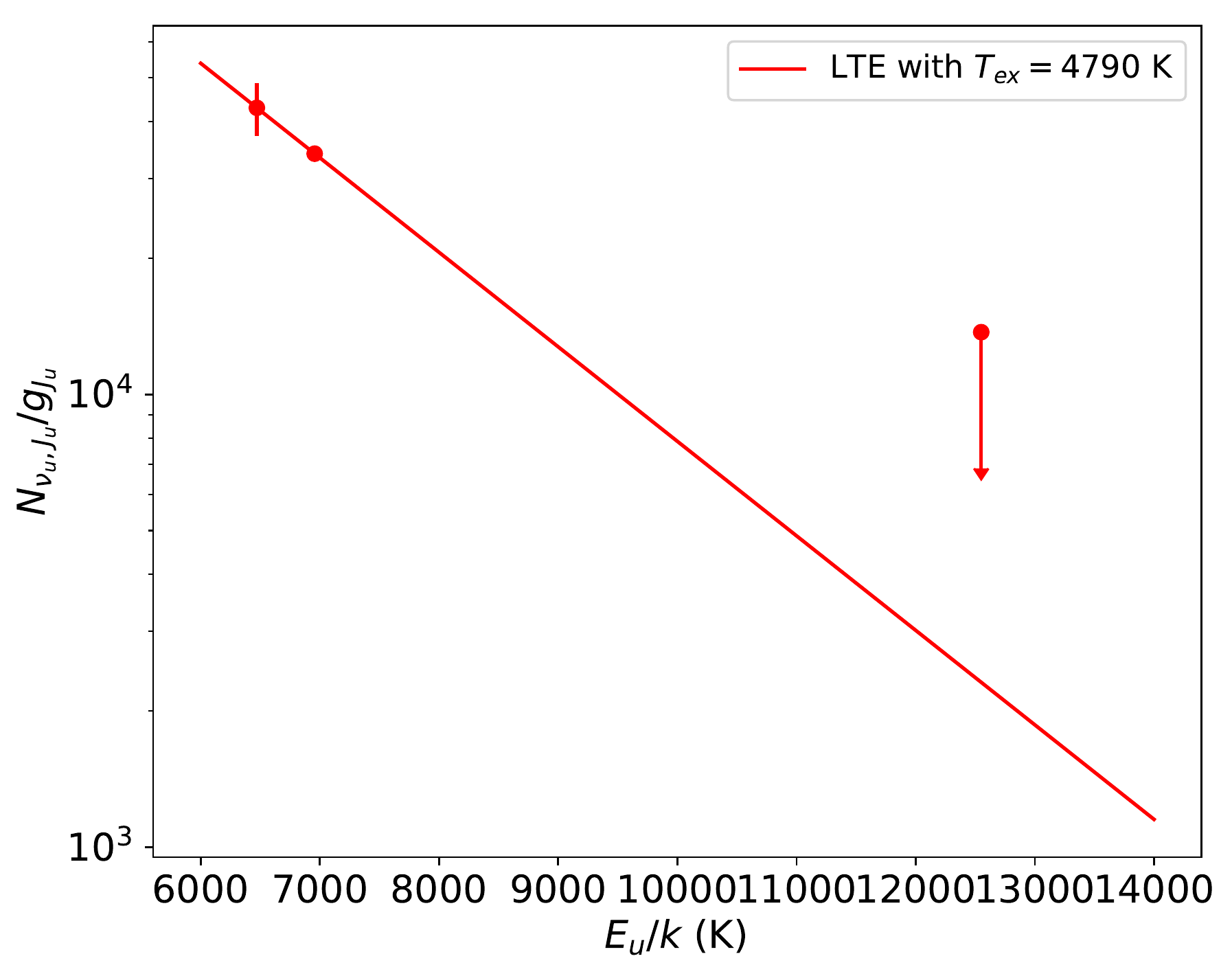}
	\caption{Excitation diagram, where we indicate the line of best fit.}
	\label{fig: excitation diagram}
\end{figure}

In Fig.~\ref{fig: excitation diagram} we show an excitation diagram, in which the strengths of the different \hh{} emission lines are used to estimate the relative populations of \hh{} molecules in different rotational ($J$) and vibrational ($\nu$) energy levels. 
Using the emission line flux $I_\text{obs}(u,l)$ corresponding to a transition between an upper $u$ and lower $l$ energy level, we plot the column density of \hh{} molecules in the upper level, $N_\text{obs}(\nu_u,J_u)$, normalised by the statistical weight $g_{J_u}$, from the expression of \citet{Rosenberg2013}:
\begin{equation}
\frac{N_\text{obs}(\nu_u,J_u)}{g_{J_u}} = \frac{4\upi\uplambda_{u,l}}{hc}\frac{I_\text{obs}(u,l)}{A(u,l)}\;,
\end{equation}
where $\uplambda_{u,l}$ is the rest-frame wavelength of the transition and $A_{u,l}$ is the spontaneous emission coefficient, here obtained from \citet{Wolniewicz1998}. 

We estimated the gas temperature by plotting $N(\nu_u,J_u)/g_{J_u}$ as a function of transition energy. In local thermodynamic equilibrium (LTE), the points will fall on a straight line in $\log N(\nu_u,J_u)/g_{J_u}$, where the slope of the line indicates the temperature. 
As we only have detections for two emission lines and an upper limit for a third, we cannot confirm whether the gas is in LTE. None the less, a linear fit to our data yields $T \approx 5000\,\rm K$.

We estimated the mass of warm \hh{} from the flux of the \hh{}\,1--0\,S(1) emission line $F_{1-0\,\rm S(1)}$ using
\begin{equation}
M_{\rm H_2} (T) = \frac{2 m_p F_{1-0\,\rm S(1)} 4 \upi D_L^2}{f_{\nu=1,J=3}(T) A_{1-0\,\rm S(1)} h \nu}
\label{eq: H2 mass}
\end{equation}
where $A_{1-0\,\rm S(1)} = 3.47\times10^{-7}\,\rm s^{-1}$ is the spontaneous emission coefficient~\citep{Turner1977}, $f_{\nu=1,J=3}(T)$ is the number fraction of \hh{} molecules in the $\nu = 1$ vibrational state and $J = 3$ rotational state at temperature $T$, and $h$ and $c$ are the Planck constant and the speed of light respectively.
In LTE, the number fraction of molecules in a rovibrational state with energy $E_j$ and degeneracy $g_j$ is described by the Boltzmann distribution
\begin{equation}
f_j(T) = \frac{g_j e^{-E_j / kT}}{Z_{\rm vr}(T)}
\end{equation}
where $k$ is the Boltzmann constant and $Z_{\rm vr}(T) = \sum_{i} g_i e^{-E_{i} / kT}$ is the partition function, which we computed using the molecular data of \citet{Dabrowski1984}.
At a temperature of $5000\,\rm K$, consistent with our excitation diagram (Fig.~\ref{fig: excitation diagram}), $f_{\nu=1,J=3}(T) = 0.0210$. 
Using this value in Eqn.~\ref{eq: H2 mass} yields $M_{\rm H_2} (5000\,\rm K) = 4400 \pm 70\,\rm M_\odot$.

\subsubsection{\feii{} emission}\label{sec: [Fe II] emission}

\begin{figure*}
	\centering
	\subcaptionbox{\label{fig: Hn4 continuum}}{\includegraphics[width=0.48\linewidth]{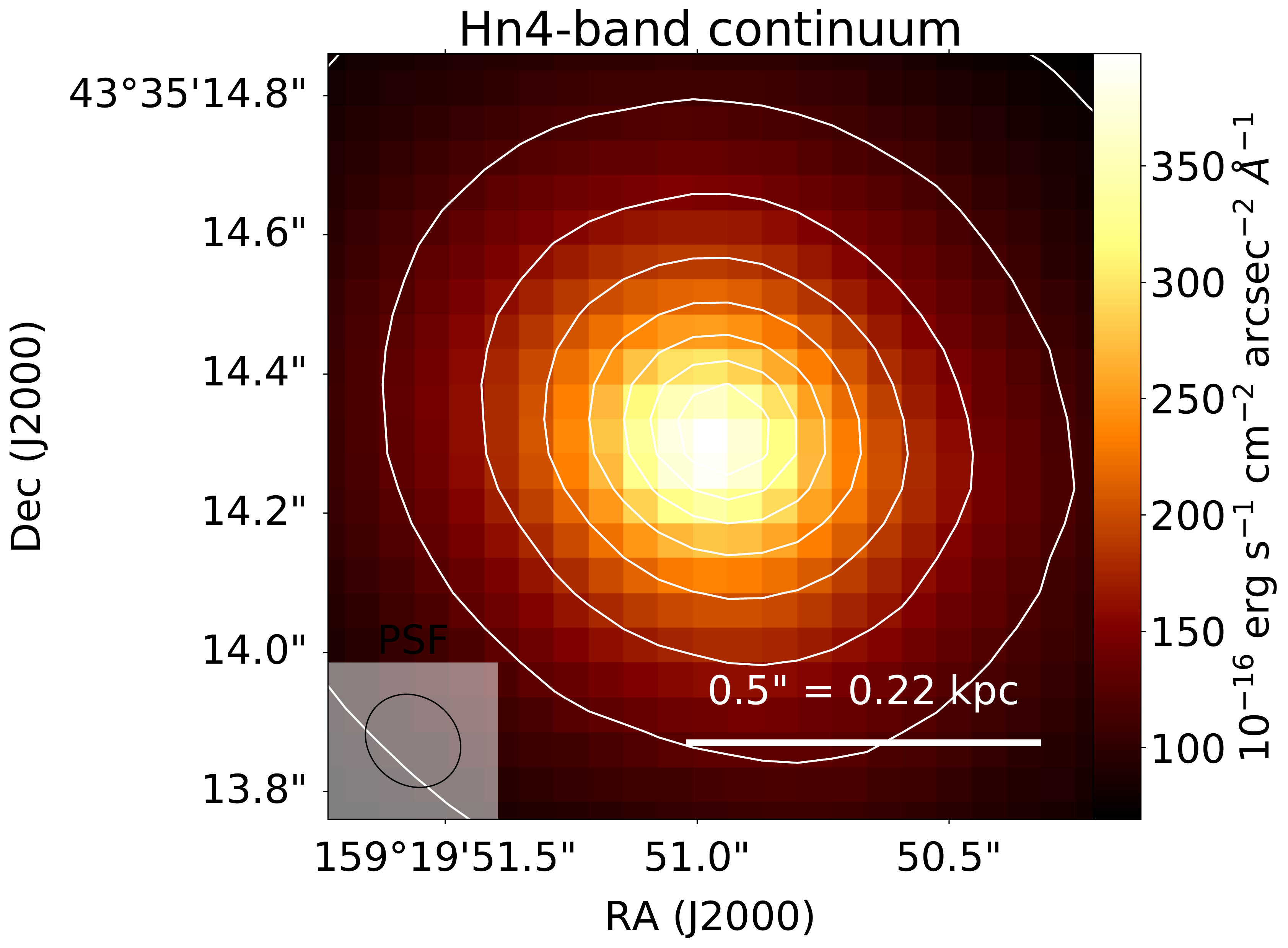}}
	\hfill
	\subcaptionbox{\label{fig: [FE II]	 flux}}{\includegraphics[width=0.49\linewidth]{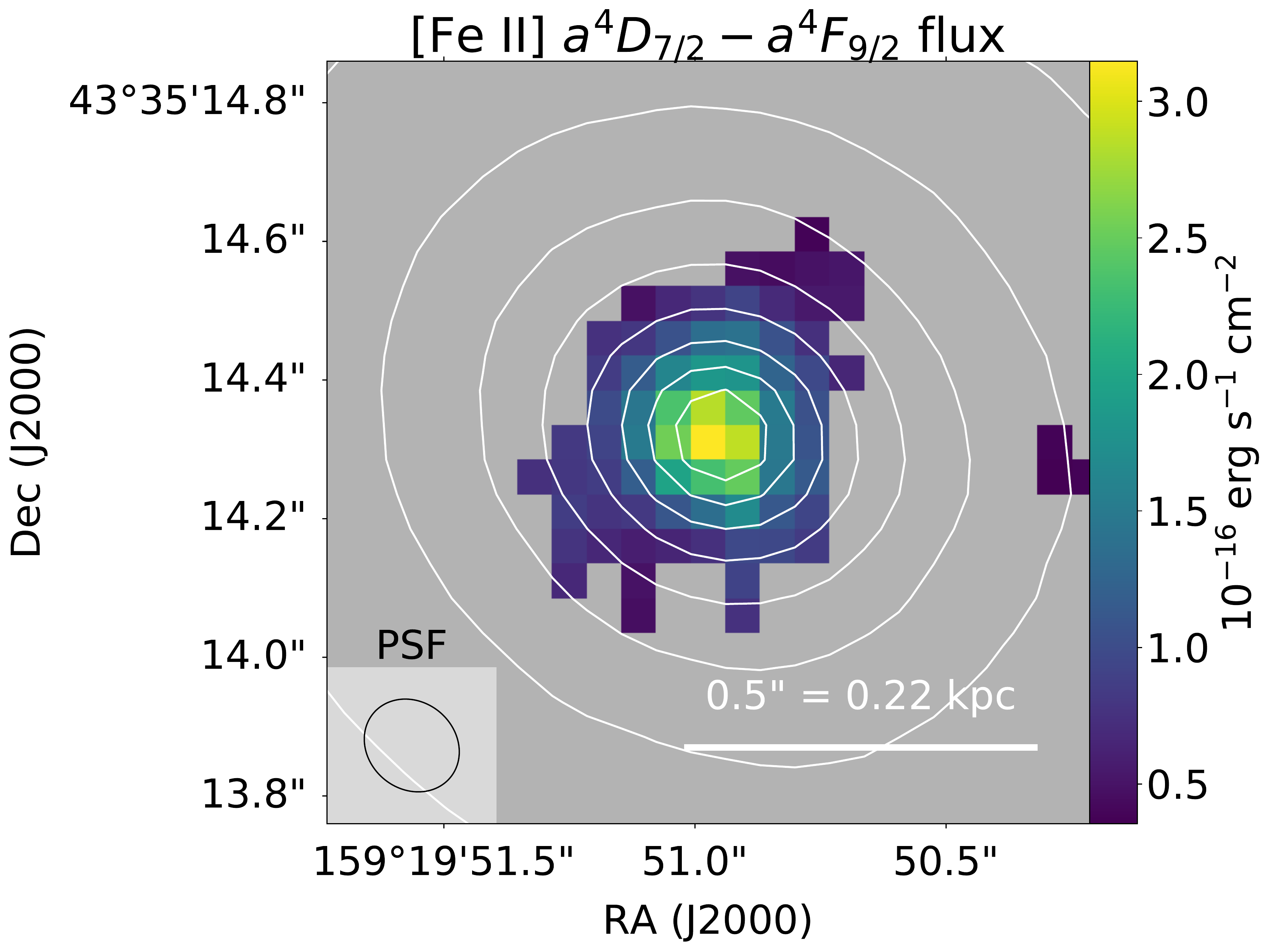}}
	\subcaptionbox{\label{fig: [FE II] vrad}}{\includegraphics[width=0.49\linewidth]{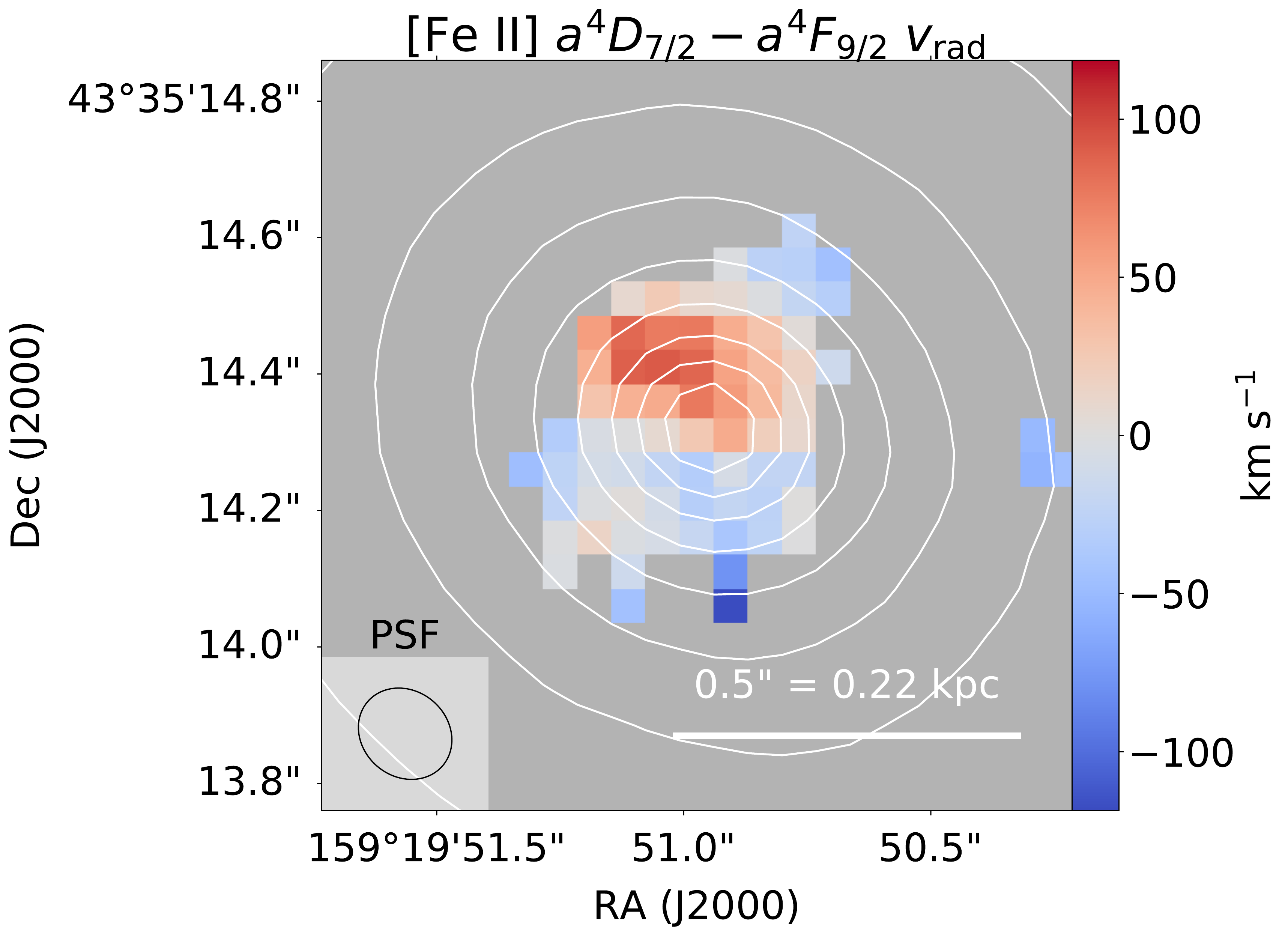}}
	\hfill
	\subcaptionbox{\label{fig: [FE II] vdisp}}{\includegraphics[width=0.49\linewidth]{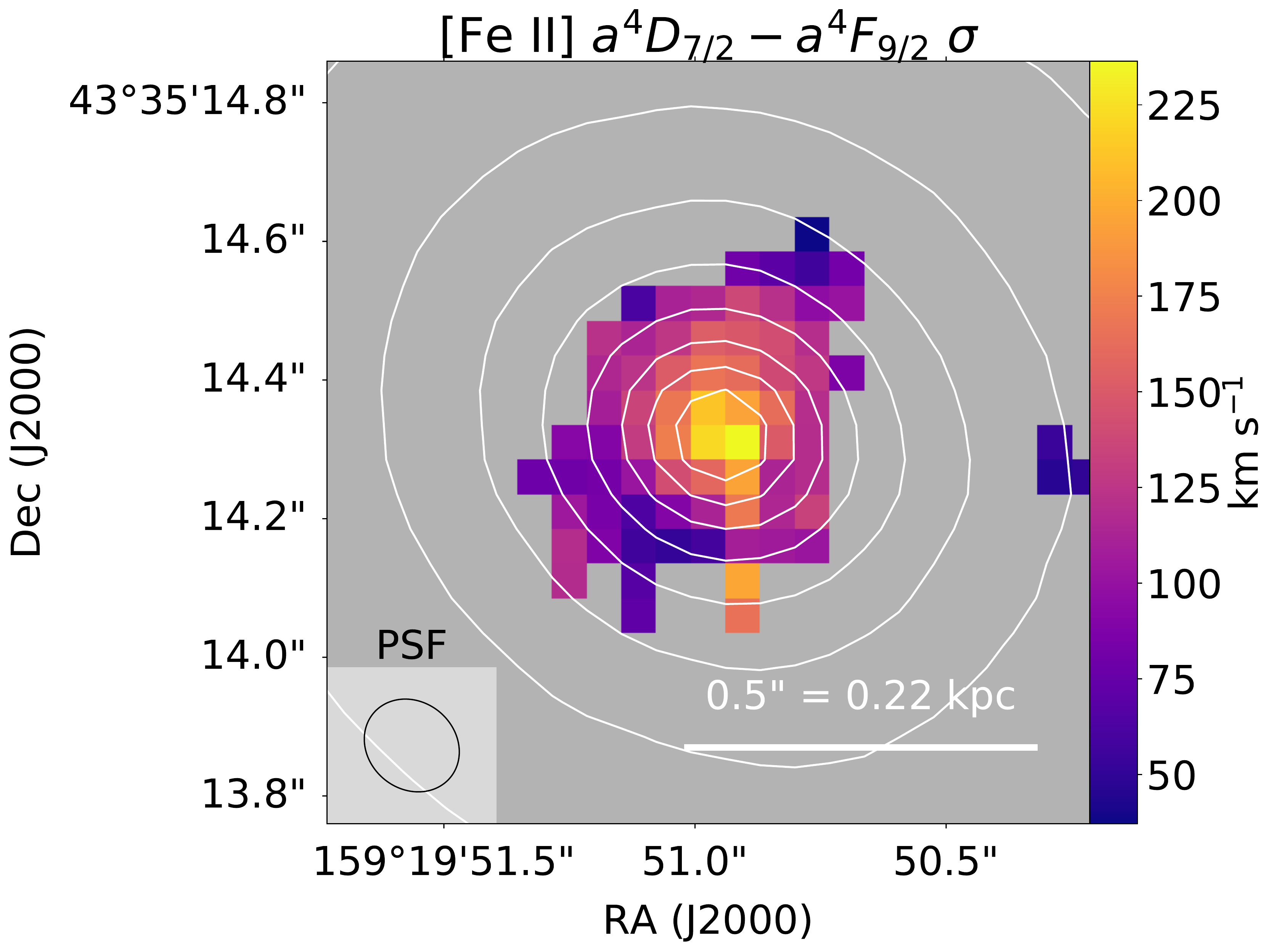}}
	\caption{(a) The Hn4\,band continuum and (b) the integrated flux, (c) radial velocity (relative to rest frame wavelength) and (d) velocity dispersion of the \feii emission line. The Hn4\,band continuum is indicated in contours, and the FWHM of the PSF (taking into account the effects of MAD smoothing) is indicated in all figures.}
	\label{fig: [Fe II] maps}
\end{figure*}

Fig.~\ref{fig: [Fe II] maps} shows the Hn4\,band continuum and the \feiilong{} emission line flux, radial velocity and velocity dispersion.

\feii{} emission occurs when dust grains are destroyed by fast shocks, releasing iron atoms into the ISM which are then singly ionised by the ambient radiation field. 
Supernovae (SNe) explosions and AGN jets or disc winds can drive shocks into the ISM, causing \feii{} emission. 

To rule out SNe explosions, we estimated the star formation rate required to produce the observed \feii{} luminosity in UGC\,05771, using the same approach as in \citet{Zovaro2019}.
First, we calculated the SNe rate $\nu_\text{SN, [Fe II]}$ from the [Fe\,\textsc{ii}]$_{1.26\,\rm \upmu m}$ luminosity using the empirical relation for SB galaxies derived by \citet{Rosenberg2012}
\begin{equation}
\log \left(\frac{\nu_{\text{SN, [Fe II]}}}{\text{yr}^{-1}}\right) = (1.01 \pm 0.2) \log \left(\frac{L({\text{[Fe\,\textsc{ii}]}_{1.26\,\rm \upmu m}})}{\text{erg s}^{-1}}\right) - (41.17 \pm 0.9)
\label{eq: SN rate from [Fe II]}
\end{equation}
where we assumed an intrinsic ratio [Fe\,\textsc{ii}] 1.26/1.64\,$\upmu$m $= 1.36$, which yielded $\nu_\text{SN, [Fe II]} = 0.28 \rm\,yr^{-1}$. 

We then estimated the SNe rate from the SFR of UGC\,05771, $\nu_{\text{SN, SFR}}$ using a solar metallicity \textsc{Starburst99}~\citep{Leitherer1999} model with a continuous SF law and a Salpeter initial mass function (IMF), where we estimated the SNe rate by scaling the SFR of the model to match the estimated \ha{}-based SFR (Section~\ref{subsubsec: Halpha SFR estimate}). For ages $> 1\,\rm Gyr$, consistent with the stellar age of the galaxy ($\sim 10\,\rm Gyr$ \citep{Sanchez2016}), this gives $\nu_{\text{SN, SFR}} \approx 0.01 \rm\,yr^{-1}$, an order of magnitude too small to cause the \feii{} emission.
Therefore, we conclude that SNe explosions are not the primary cause of the \feii{} emission.

\section{CALIFA observations}\label{sec: CALIFA data}

\subsection{Observations and data reduction using \textsc{Pipe3D}}\label{subsec: CALIFA observations}

\begin{figure*}
	\centering
	\includegraphics[width=1\linewidth]{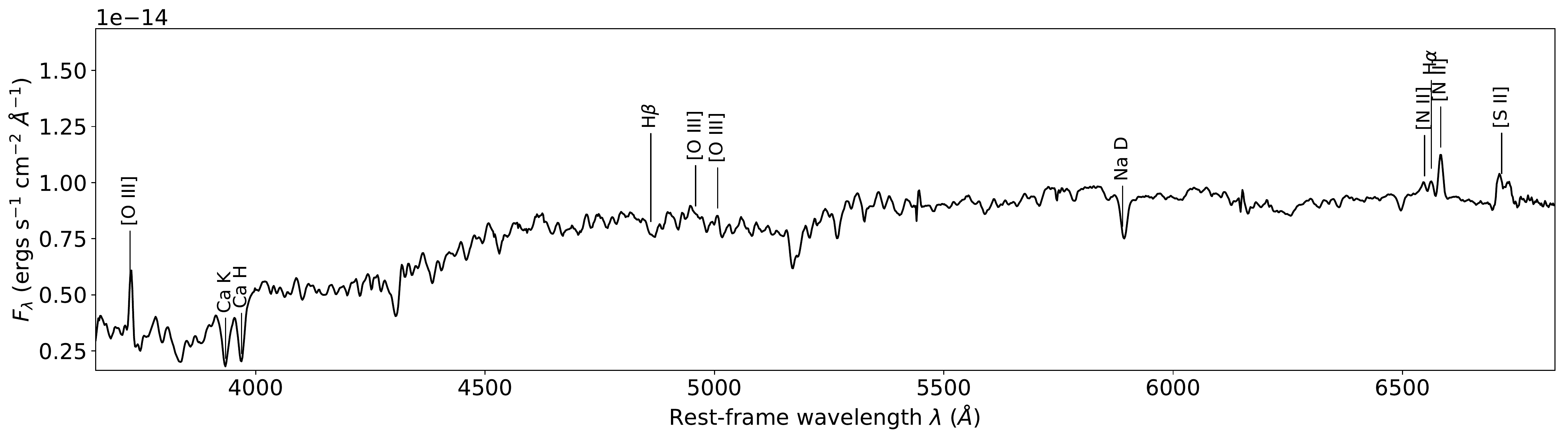}
	\caption{Integrated spectrum extracted from the CALIFA data cube from spaxels within 5\,kpc of the nucleus.}
	\label{fig: CALIFA spectrum}
\end{figure*}

UGC\,05771 was observed using the Potsdam Multi Aperture Spectrograph (PMAS) instrument~\citep{Roth2005} on the 3.5 m telescope at the Calar Alto observatory as a part of the Calar Alto Legacy Integral Field Area (CALIFA) survey of $0.005 < z < 0.03$ galaxies~\citep{Sanchez2012,Walcher2014,Sanchez2016}. 
PMAS is a fibre-bundle spectrograph that produces a hexagonal data cube with 2.5'' spatial resolution (FWHM) over a 74'' $\times$ 64'' field of view. Each galaxy in the survey was observed in both high ($R\sim1650$) and low spectral resolution ($R\sim850$) modes of PMAS, covering wavelength ranges of $3700$ to $4800$\,\AA{} and $3749$ to $7500$\,\AA{} respectively. The spectral resolution of the high- and low-resolution modes corresponds to velocity dispersions of approximately $100\,\textrm{km\,s}^{-1} < \sigma < 200\,\textrm{km\,s}^{-1}$ and $60\,\textrm{km\,s}^{-1} < \sigma < 80\,\textrm{km\,s}^{-1}$ respectively.
We show the spectrum extracted from the `combined' data cube, which merges the high- and low-resolution spectra into a single data cube, from spaxels within 5\,kpc of the nucleus in Fig.\ref{fig: CALIFA spectrum}.

We used the data products produced by \textsc{Pipe3D}, a processing pipeline developed for integral field unit surveys~\citep{Sanchez2016} developed by the CALIFA collaboration, to study the spatially resolved stellar population and ionised gas properties of UGC\,05771.
To elevate the S/N to a level sufficient for analysis of the stellar continuum, \textsc{Pipe3D} spatially bins the data cube, before fitting spectral templates to the spectra to provide quantities including the kinematics, metallicity and ages of the stellar population. 
The stellar continuum fit to each spaxel is then subtracted from the data cube, yielding an emission line-only data cube, to enable analysis of the emission line spectra.

\subsection{Stellar properties}\label{subsec: CALIFA stellar properties}

\begin{figure*}
	\centering
	\subcaptionbox{\label{fig: stellar vrad}}{\includegraphics[width=0.49\linewidth]{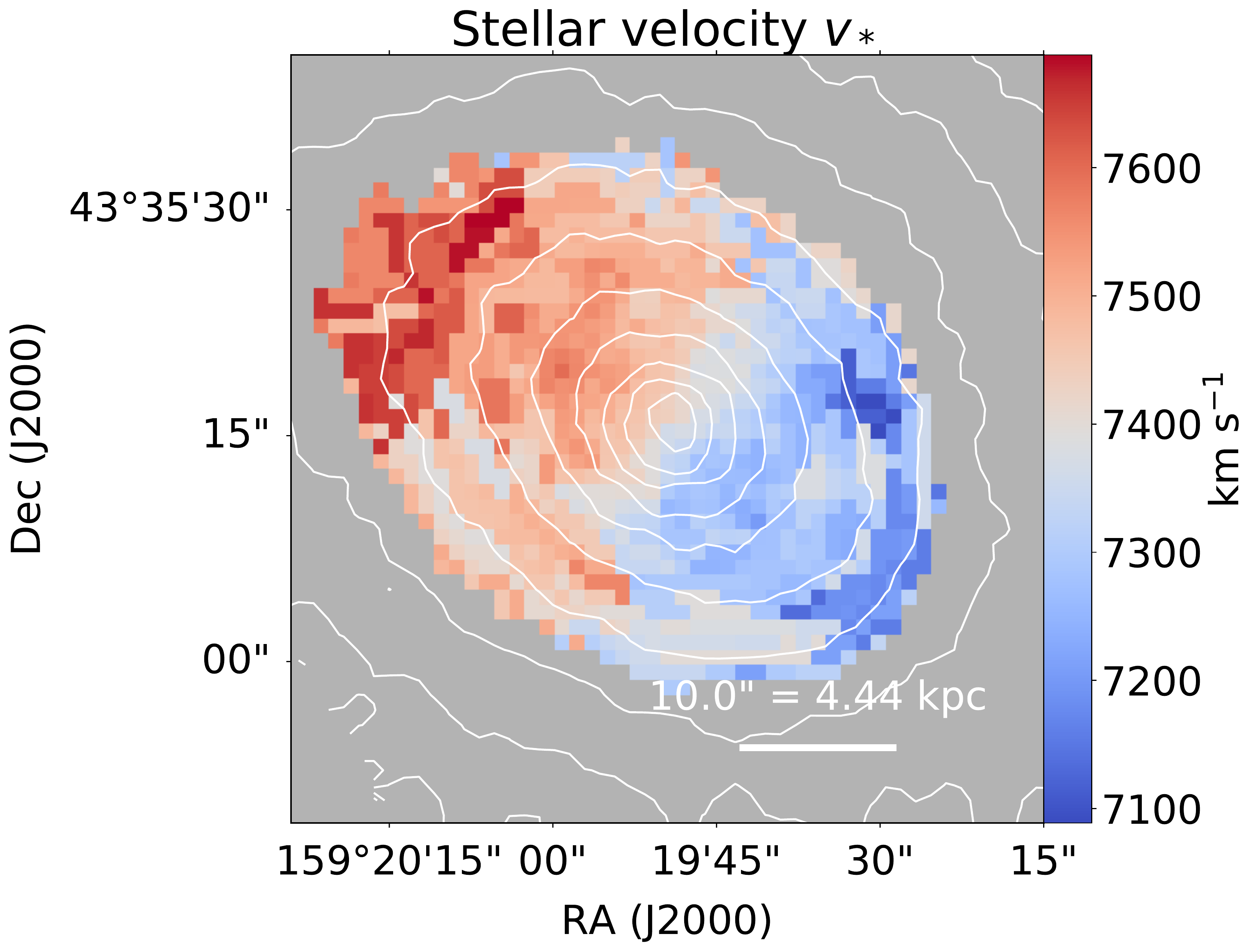}}
	\hfill
	\subcaptionbox{\label{fig: stellar vdisp}}{\includegraphics[width=0.49\linewidth]{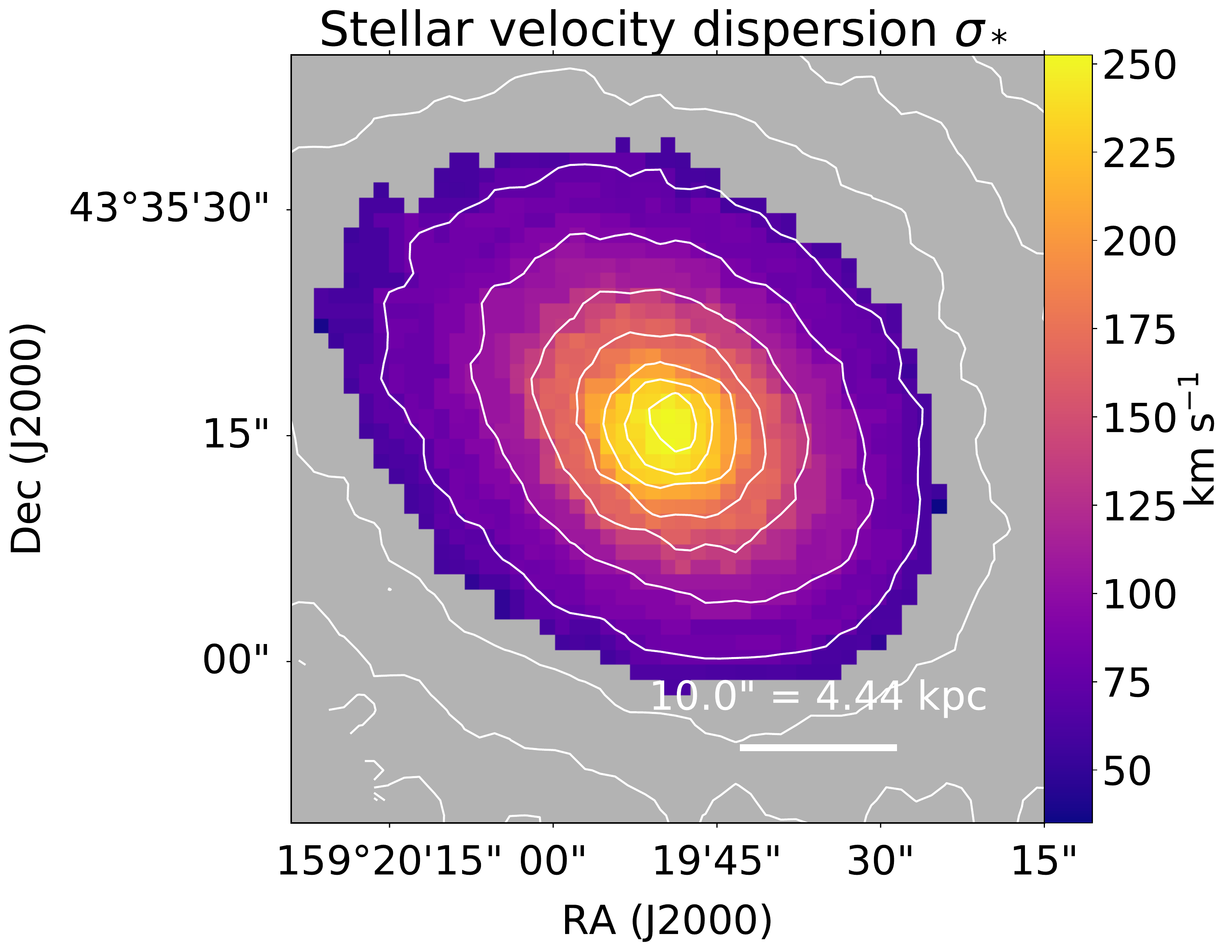}}
	\caption{(a) The stellar radial velocity, relative to systemic, and (b) velocity dispersion (Gaussian $\sigma$). The contours indicate the logarithmically-scaled $V$\,band continuum.}
	\label{fig: stellar kinematics}
\end{figure*}

The best-fit stellar population of UGC\,05771 produced by \textsc{Pipe3D} is predominantly old, with a star formation history well-approximated by an instantaneous burst $\sim 10\,\rm Gyr$ ago. 
Fig.~\ref{fig: stellar kinematics} shows the $V$\,band continuum reconstructed using the data cube, the stellar radial velocity and velocity dispersion.

We estimated the black hole mass using the $M_{\rm BH}-\sigma_*$ relation of \citet{Gultekin2009}, given by
\begin{multline}
\log_{10} \left[ \frac{M_{\rm BH}}{\textrm{M}_\odot} \right] = \left(8.23 \pm 0.08\right) \pm \left(3.96 \pm 0.42\right) \\
\log_{10} \left[\frac{\sigma_e}{200 \rm \,km\,s^{-1}}\right]
\end{multline}
where $\sigma_e$ is the luminosity-weighted stellar velocity dispersion within the effective radius $R_e$.
We calculated $\sigma_e$ from the average flux intensity $I_i$ and stellar velocity dispersion $\sigma_{*,i}$ in each spaxel $i$ using
\begin{equation}
\sigma_e = \frac{\sum_i I_i \sigma_{*,i}}{\sum_i \sigma_{*,i}}.
\end{equation}
We calculated $\sigma_e$ within 1 $R_e$, which we estimated by fitting a S\'ersic profile to the $V$\,band continuum image constructed from the data cube, which yields $\sigma_{*,i} = 226 \pm 3\,\rm km\,s^{-1}$.
Using this value, we estimated $\log_{10} M_{\rm BH} = 8.54 \pm^{0.23}_{0.03} \log \rm\,M_\odot$. 

\begin{figure*}
	\centering
	\subcaptionbox{\label{fig: CALIFA V-band continuum}}{\includegraphics[width=0.49\linewidth]{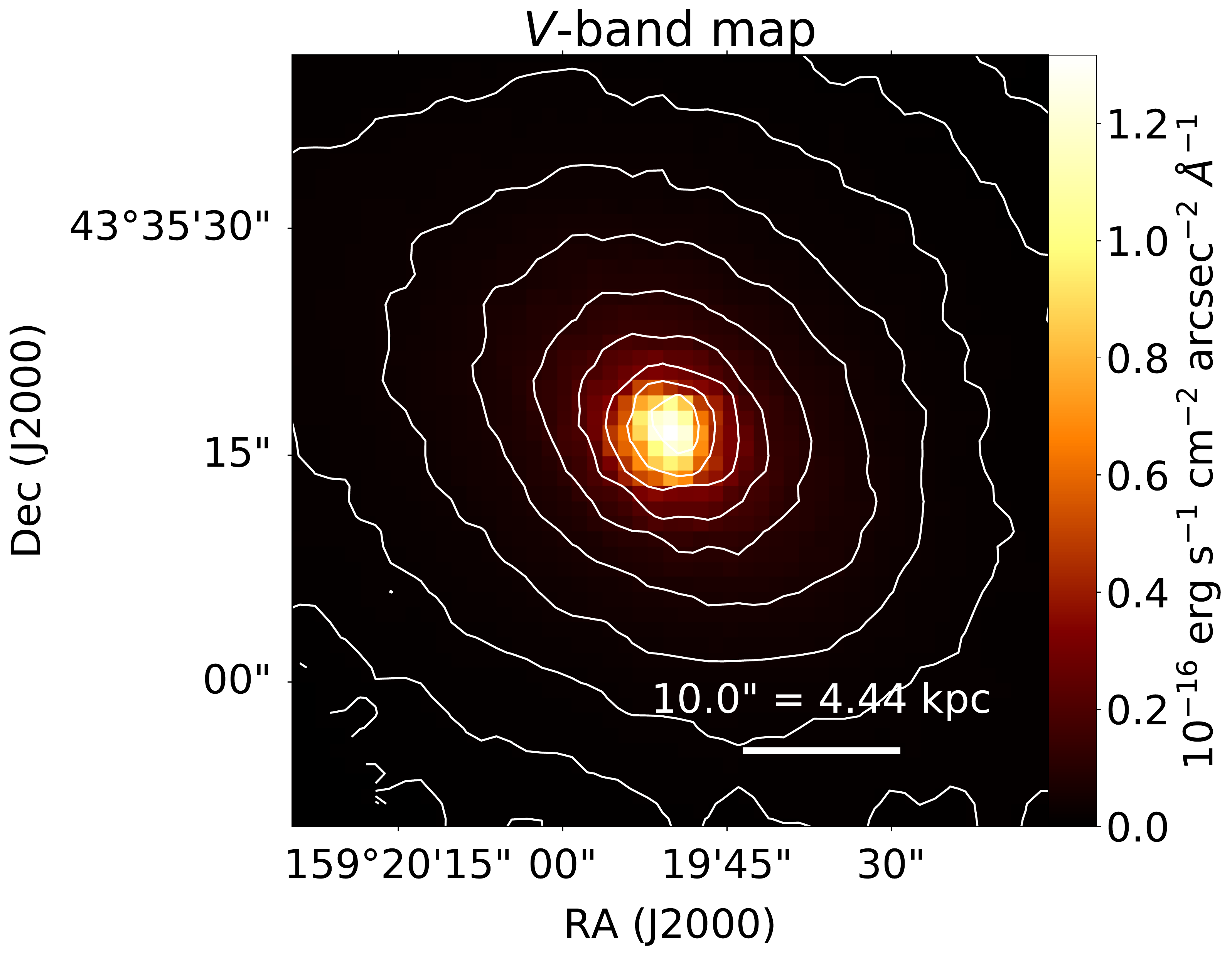}}
	\hfill
	\subcaptionbox{\label{fig: Halpha flux}}{\includegraphics[width=0.49\linewidth]{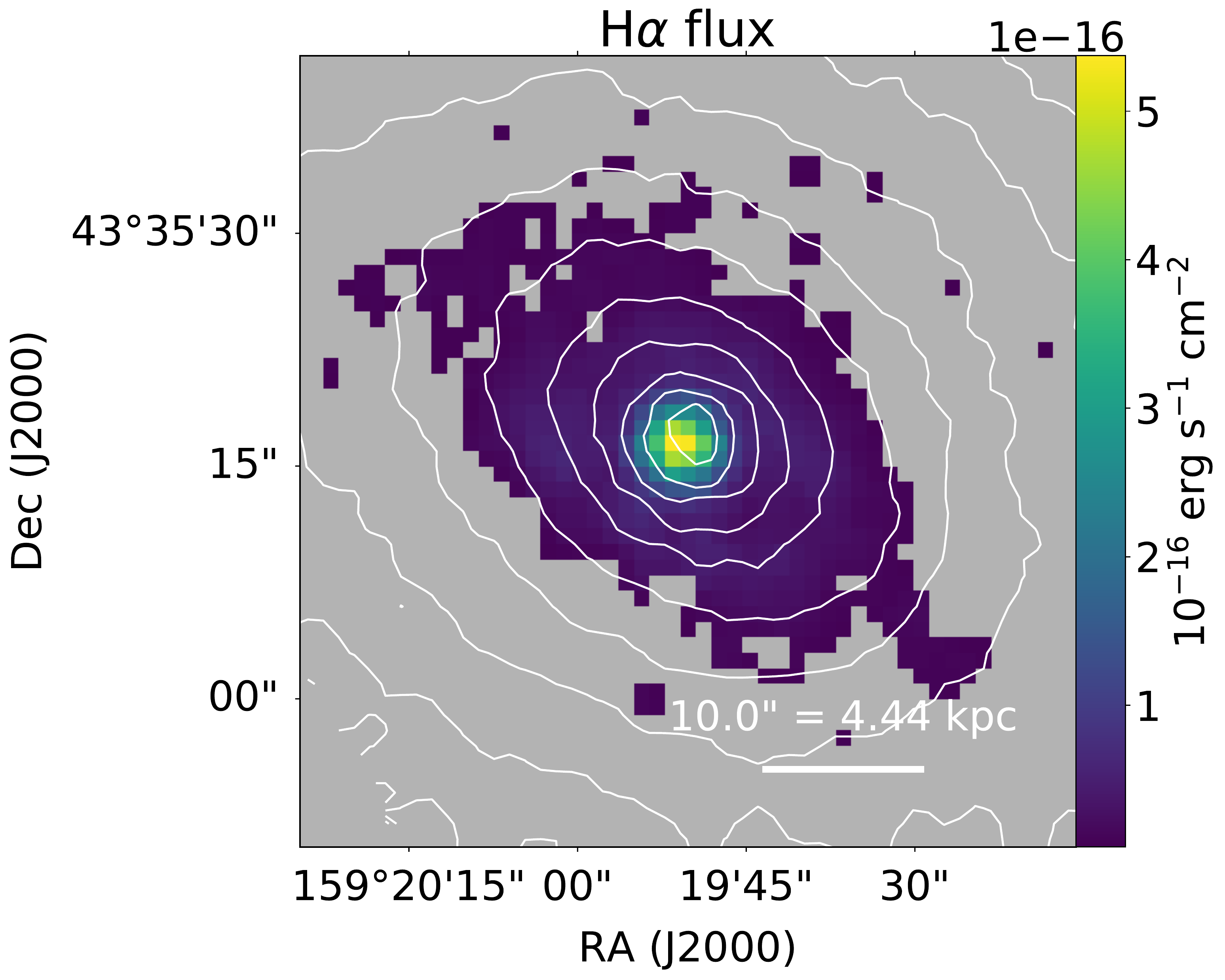}}
	\subcaptionbox{\label{fig: Halpha vrad}}{\includegraphics[width=0.49\linewidth]{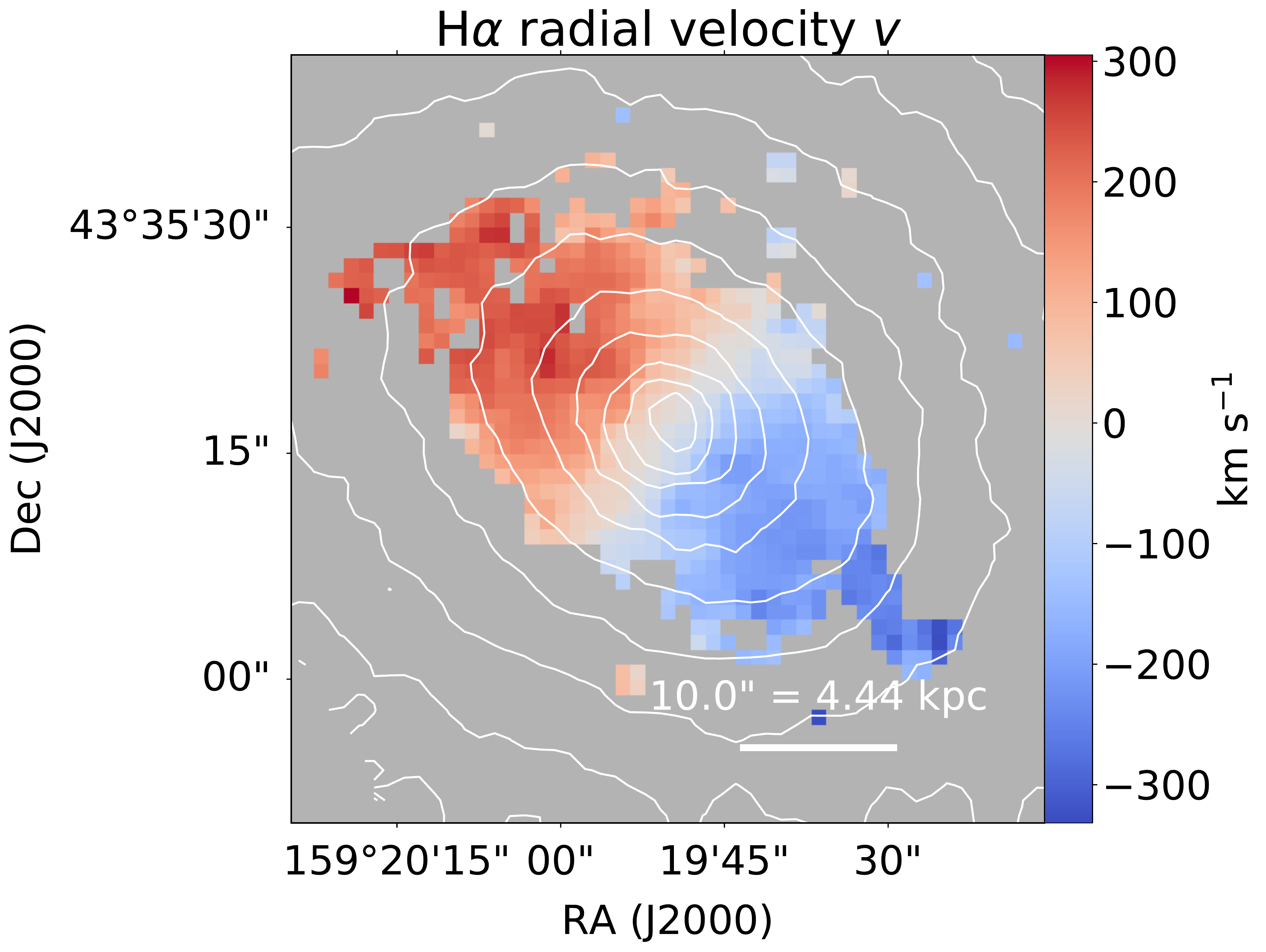}}
	\hfill
	\subcaptionbox{\label{fig: Halpha vdisp}}{\includegraphics[width=0.49\linewidth]{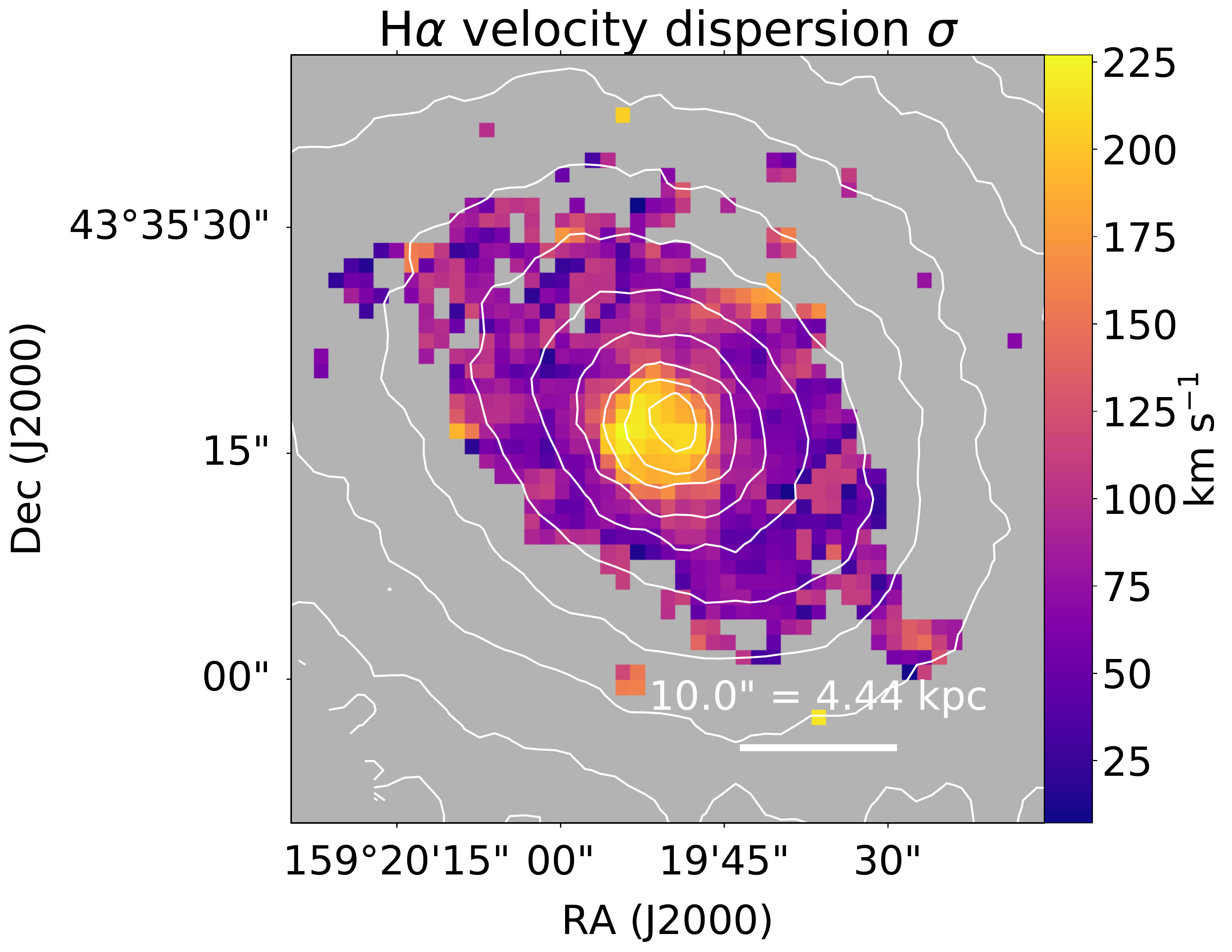}}
	\caption{(a) $V$\,band continuum; (b) \ha{} emission line flux, (c) radial velocity (relative to systemic) and (d) velocity dispersion (Gaussian $\sigma$), corrected for the instrumental dispersion. Contours show the logarithmically-scaled $V$\,band continuum.}
	\label{fig: Halpha flux, vel and vdisp}
\end{figure*}

\subsection{Emission line analysis}\label{subsec: CALIFA emission line analysis}

We performed our own Gaussian emission line fits to the emission line-only data cube using the method detailed in Section~\ref{subsec: OSIRIS Emission line fitting}. We simultaneously fitted the emission lines {[}O\,\textsc{ii}{]}$\uplambda\uplambda$3726,3729, \hb{}, {[}O\,\textsc{iii}{]}$\uplambda\uplambda$4959,5007, {[}N\,\textsc{ii}{]}$\uplambda\uplambda$6548,6583, \ha{}, {[}S\,\textsc{ii}{]}$\uplambda$6716 and {[}S\,\textsc{ii}{]}$\uplambda$6731, where we constrained each line to have the same kinematics, and we fixed the relative fluxes of the lines in the [N\,\textsc{ii}] and [O\,\textsc{iii}] doublets to their expected values of 1:3.06 and 1:2.94 respectively. Due to the low spectral resolution of PMAS at the wavelengths of most emission lines, we only fitted a single kinematic component to each line.	

We show the total extinction-corrected emission line fluxes in Table~\ref{tab: eline fluxes}, where we have summed fluxes from spaxels with fits that have $\rm S/N > 3$ and reduced $\chi^2 < 2$. 
The ionised gas kinematics are shown in Fig.~\ref{fig: Halpha flux, vel and vdisp}, where we present the \ha{} flux, radial velocity and velocity dispersion.

\subsubsection{Reddening}\label{subsubsec: CALIFA reddening}

\begin{figure}
	\centering
	\includegraphics[width=1\linewidth]{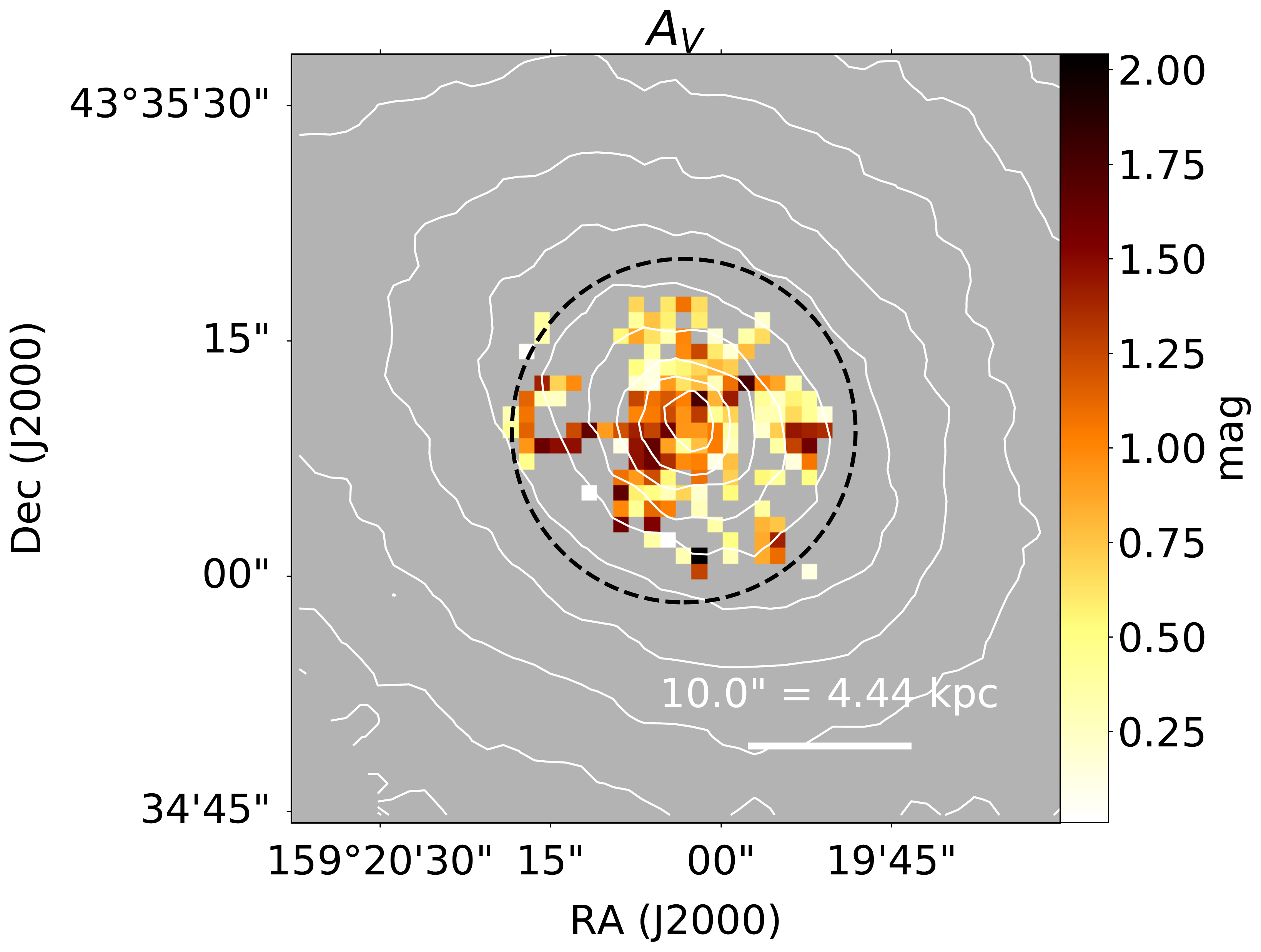}
	\caption{Total extinction in the $V$ band $A_V$. The contours indicate the logarithmically-scaled $V$\,band continuum, and the dashed circle shows the FWHM of the IRAM 30\,m telescope beam at the observed frequency of the CO(1--0) line (Section~\ref{sec: IRAM data}).}
	\label{fig: A_V map}
\end{figure}

To correct our emission line fluxes for extinction, we used the reddening curve of \citet{Fitzpatrick&Massa2007} with $R_V = 3.1$, assuming that the ratio of the intrinsic line fluxes $I(\textrm{H}\alpha)/I(\textrm{H}\beta) = 2.85$ corresponding to Case B recombination. 
We calculated the total extinction in the $V$\,band, $A_V$, in each spaxel in which the line ratio had a $\rm S/N > 3$ (Fig.~\ref{fig: A_V map}). 
There were a number of spaxels in which the ratio of the observed fluxes $\frac{F(\textrm{H}\alpha)}{F(\textrm{H}\beta)} < 2.85$, corresponding to values of $A_V < 0$; these non-physical line ratios are most likely due to systematic errors in fitting the underlying stellar continuum. 

To estimate the mean $A_V$, we calculated the average of the \ha{}/\hb{} ratios in each spaxel in which the ratio has $\rm S/N > 3$ and $\frac{F(\textrm{H}\alpha)}{F(\textrm{H}\beta)} > 2.85$, which yielded $A_V = 0.831 \pm 0.014$.

In spaxels where the \hb{} $\rm S/N < 3$ but \ha{} is still detected, we estimated upper limits for the \hb{} flux using the method detailed in Section~\ref{subsec: OSIRIS Emission line fitting}, which provided a lower limit for $A_V$. In spaxels closer to the nucleus, $A_V \gtrsim 0.5$; at larger radii, where the \ha{} surface brightness drops off, we cannot provide any meaningful constraint on $A_V$. 
Therefore, without any further information, we assumed $A_V = 0.831 \pm 0.014$ across the whole galaxy, and have used this value to correct for extinction in the line fluxes given in Table~\ref{tab: eline fluxes}.

We also used $A_V$ to estimate the hydrogen column density $N_{\rm H}$ using the relation of \citet{Gulver&Ozel2009}. In the spaxels over which we could measure $A_V$, the mean column density $N_{\rm H} = 1.84 \pm 0.03 \times 10^{21}\,\rm cm^{-2}$.
We estimated the total hydrogen mass from the column densities in each spaxel $i$ using $M_{\rm H} = \sum_i \mu m_{\rm H} N_{\rm H,\,i} A_i \geq 4.7 \pm 0.2 \times 10^8 \, \rm M_\odot$ where $\mu = 1.37$ is the mean molecular mass, $m_{\rm H}$ is the mass of the hydrogen atom and $A_i$ is the area of each spaxel in $\rm cm^{2}$. 
This gave a mean molecular gas mass surface density $\Sigma_{\rm gas} = 10.0 \pm 0.5 \, \rm M_\odot \,\,pc^{-2}$ which is consistent with our estimate based on our CO observations ($\Sigma_{\rm gas} = 15 \pm 5 \, \rm M_\odot \,\,pc^{-2}$; Section~\ref{sec: IRAM data}).

\subsubsection{Excitation mechanism}\label{subsubsec: ODDs}

To investigate the excitation source of the ionised gas, we constructed optical diagnostic diagrams (ODDs)~\citep{Baldwin1981,Veilleux&Osterbrock1987,Kewley2001,Kauffmann2003,Kewley2006}. 
In an ODD, the ratios of certain pairs of emission line fluxes are plotted against one another to distinguish between excitation due to shocks, photoionisation from stellar radiation fields and harder radiation fields from AGN. 

\begin{figure*}
	\centering
	\subcaptionbox{\label{fig: BPT_dist}}{		
		\includegraphics[width=1\textwidth]{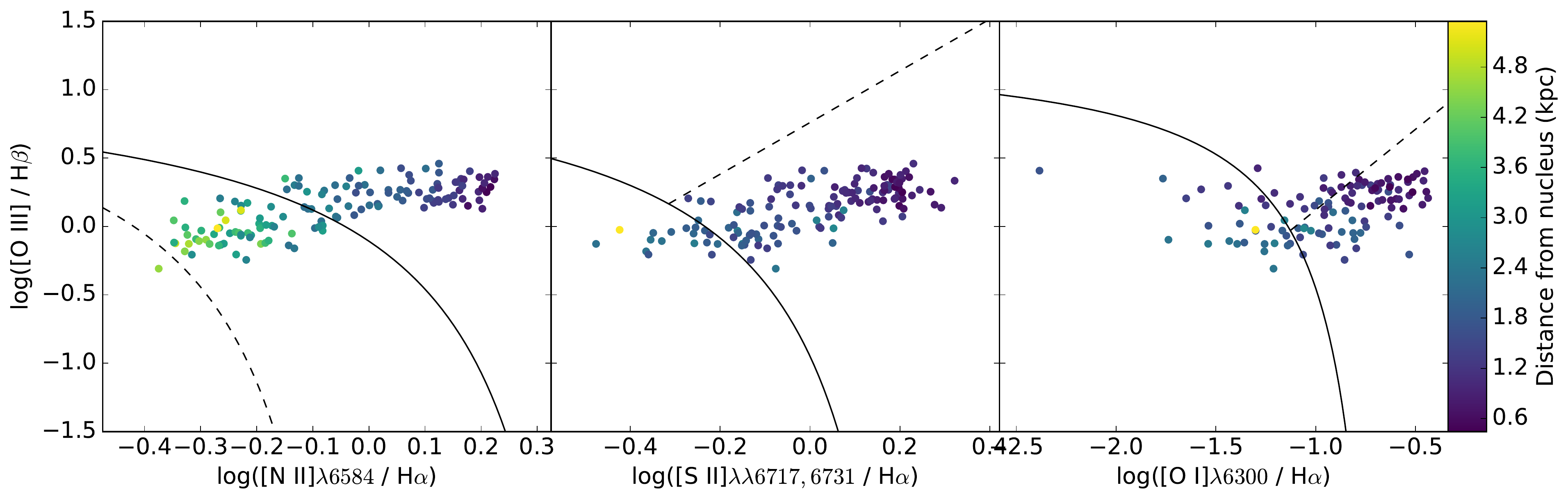}
	}
	\hfill
	\subcaptionbox{\label{fig: BPT_vdisp}}{		
		\includegraphics[width=1\textwidth]{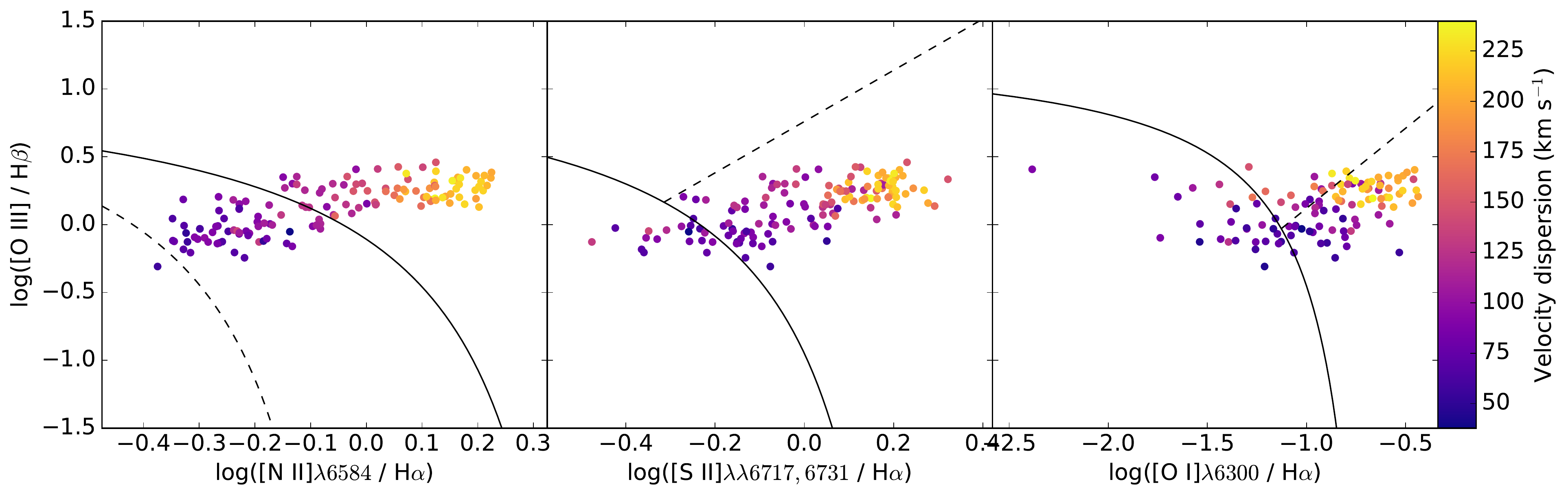}
	}
	\caption{
	Optical diagnostic diagrams for spaxels in UGC\,05771~\protect\citep{Baldwin1981,Veilleux&Osterbrock1987} with points coloured by (a) their distance from the nucleus (assumed to coincide with the peak in the $V$\,band continuum) and (b) by the \ha{} velocity dispersion. 
	The solid lines represent the maximum [O\,\textsc{iii}]/\hb{} ratio that can arise from star formation alone, derived from photoionisation models~\protect\citep{Kewley2001}. 
	In the left panels, the dashed line is the equivalent empirical relation of \protect\citet{Kauffmann2003} which separates star-forming galaxies and AGN hosts. 
	In the middle and right panels, the dashed lines separate Seyfert-like (above) and LINER-like ratios (below the line)~\protect\citep{Kewley2006}.
}
	\label{fig: ODDs}
\end{figure*}

In Fig.~\ref{fig: BPT_dist}, we show the spaxel-by-spaxel ODDs for UGC\,05771, where each spaxel is colour-coded by its projected distance from the nucleus in the plane of the sky. We used the line fluxes and velocity dispersion from our Gaussian fit, only including spaxels in which the S/N of all lines exceeds 3, except for the [O\,\textsc{i}]$\uplambda 6300$ line for which we used the flux from the Monte Carlo-method fit provided by \textsc{Pipe3D}~\citep[][\S 3.6]{Sanchez2016} because the S/N in this line was too low to be fit using our own method.
In all 3 ODDs, most spaxels lie above the maximum [O\,\textsc{iii}]/\hb{} ratio that can arise from star formation alone at a given [N\,\textsc{ii}]/\ha{} ratio (the solid curve) of~\cite{Kewley2001}. 

Despite their diagnostic power, line ratios alone cannot be used to distinguish between photoionisation and shock excitation for spaxels with low-ionization nuclear emission line region (LINER)-like emission, which otherwise occupy overlapping regions to the right of all 3 ODDs. 
For this reason, in Fig.~\ref{fig: BPT_vdisp} we show the same ODDs with the spaxels colour coded by the \ha{} velocity dispersion, which reveals that the spaxels with the largest [N\,\textsc{ii}]/\ha{} ratios are also those with the highest velocity dispersion, indicating shocks dominate the line emission.

Although we cannot rule out shocks as the excitation mechanism for the line emission in the outer regions of the galaxy, the lower velocity dispersion in these regions is consistent with line emission powered by a diffuse ionisation field, probably from evolved stars~\citep{Singh2013,Belfiore2016}. This is consistent with the $\sim 10\,\rm Gyr$ old stellar population of UGC\,05771, in which post-AGB stars, hot and evolved stars that emit a hard ionising spectrum, are expected to dominate the ionisation field, contaminating SFR estimates based on the \ha{} flux and FUV magnitude. We discuss the implications of this in the following section.

\subsubsection{Star formation rate estimate}\label{subsubsec: Halpha SFR estimate}

Using the total extinction-corrected \ha{} flux and the calibration of \citet{Calzetti2013} derived from stellar population modelling using a Kroupa IMF, we estimated the global $\rm SFR(H\upalpha) = 0.2795 \pm 0.0011 \, M_\odot\,yr^{-1}$.
However, we are confident this is an upper limit due to contamination from other sources of \ha{} emission.
Inspection of Fig.~\ref{fig: Halpha flux} reveals that the \ha{} emission peaks sharply within a few\,kpc of the nucleus, which is also where we observe very strong shock signatures (Fig.~\ref{fig: ODDs}). Therefore a substantial portion of the total \ha{} flux is due to shocks. 

To obtain a more strict upper limit to the SFR, we used the ODDs presented in Section~\ref{subsubsec: ODDs}.
Although many spaxels lie in the LINER region of the ODD, a significant number of spaxels occupy the intermediate region (between the dashed and solid curves in the left panels of Figs.~\ref{fig: BPT_dist} and \ref{fig: BPT_vdisp}), in which the line ratios indicate that both star formation and shocks and/or radiation from post-AGB stars contribute to the line emission.
To obtain a better upper limit for the SFR, we only included the \ha{} flux from spaxels that lie in the intermediate and star-forming regions.	
Unfortunately, there are many spaxels that we could not show on the ODDs due to poor S/N in [O\,\textsc{iii}] and \hb{}. However, these spaxels are predominantly at large radii, where the surface brightness is low, and inspection of \ref{fig: BPT_dist} shows that there are very few spaxels at radii $\gtrsim 2 \,\rm\,kpc$ that lie in the LINER region. 
Hence, we assumed that all spaxels at large radii with unknown line ratios lie in the intermediate or star-forming regions of the ODD. 
Summing the \ha{} fluxes from all spaxels in the intermediate or star-forming regions yields $\rm SFR(\rm H\upalpha) = 0.1046 \pm 0.0007 \rm \, M_\odot \, yr^{-1}$. We stress that this again represents an upper limit, as the majority of these spaxels lie in the intermediate region, implying contributions from sources other than star formation. 

To determine whether there is any significant star formation obscured by extinction, we also estimated the SFR using far-IR fluxes from the \textit{Infrared Astronomical Satellite}~\citep[\textit{IRAS},][]{Neugebauer1984}. Because UGC\,05771 was not detected by \textit{IRAS}, we instead used the \textit{IRAS} Scan Processing and Integration tool\footnote{Available \url{https://irsa.ipac.caltech.edu/applications/Scanpi/}.} to estimate upper limits from the total integrated flux density within $\pm 2.5'$ of UGC\,05771. Using the 60\,$\upmu$m flux density upper limit of $f_\nu(60 \,\rm \upmu m) = 0.12 \, \rm Jy$ and the total IR (TIR) bolometric correction $L_{\rm TIR} = 1.7 L_{60 \, \upmu \rm m}$ of \citet{Rowan-Robinson1997} and the TIR SFR calibration of \citet{Kennicutt1998}, we obtained $\rm SFR(TIR) < 0.5 \, \rm M_\odot \, yr^{-1}$. Given that the FIR fluxes may also be contaminated by the AGN, this confirms that there is unlikely to be a significant amount of obscured star formation, which is expected given the modest $A_V$ we estimate from our \ha{} and \hb{} fluxes.

We also estimated the SFR using the \textit{Galaxy Evolution Explorer} far-UV magnitude.
After correcting for reddening, and using the UV SFR law of \citet{Salim2007}, we found $\textrm{SFR}(\textrm{UV}) = 0.94 \pm 0.12\,\rm \rm \, M_\odot \, yr^{-1}$, much larger than that based on the \ha{} and far-IR fluxes. However, the UV flux is likely to be strongly contaminated by emission from old stars; as discussed above, LINER-like emission line ratios suggest that post-AGB stars most likely dominate the UV radiation field in the outer regions of the galaxy. Hence, $\textrm{SFR}(\textrm{UV})$ represents a strict upper limit.

Therefore, because $\textrm{SFR}(\textrm{TIR})$, $\textrm{SFR}(\textrm{UV})$ and $\rm SFR(\rm H\upalpha)$ represent strict upper limits, we conclude that the true SFR in UGC\,05771 is most likely lower than $\rm SFR(\rm H\upalpha) = 0.1046 \pm 0.0007 \rm \, M_\odot \, yr^{-1}$. 
We use this latter estimate of the SFR to investigate the star formation efficiency in UGC\,05771 in Section~\ref{sec: Conclusion}.

\subsubsection{Ionised gas kinematics}\label{subsubsec: CALIFA ionised gas kinematics}

Fig.~\ref{fig: Halpha flux, vel and vdisp} shows the \ha{} flux, radial velocity and velocity dispersion.
Upon first inspection, there are no obvious signs of disrupted kinematics in the \ha{} radial velocity. 
This is perhaps to be expected, given the low radio luminosity of the source, which is several orders of magnitude lower than those of the sample of GPS and CSS sources of \citet{Holt2008} in which outflows of up to a few $1000\,\rm km\,s^{-1}$ are observed. 

None the less, to determine whether there are any significant non-circular motions in the ionised gas, we fitted a disc model to the radial velocity data using \textsc{mpfit}.
We first fitted a S\'ersic profile to the galaxy's light profile along its semi-major axis in the $V$\,band continuum (Fig.~\ref{fig: CALIFA V-band continuum}); we show the S\'ersic parameters in Table~\ref{tab: Properties of UGC05771}. 
We then used the analytical expressions of \citet{Terzic&Graham2005} for the circular velocity in a S\'ersic potential, which also requires the total galaxy mass; we used the mass estimated by \citet{Sanchez2016} (Table~\ref{tab: Properties of UGC05771}).
We fitted the circular velocity profile to our data, allowing the systemic velocity, inclination and position angle to vary.
We constrained the kinematic centre of the disc to coincide with the peak in the $V$\,band continuum.

\begin{figure*}
	\centering
	\subcaptionbox{\label{fig: Halpha_vmeas Gaussian}}{\includegraphics[width=0.49\linewidth]{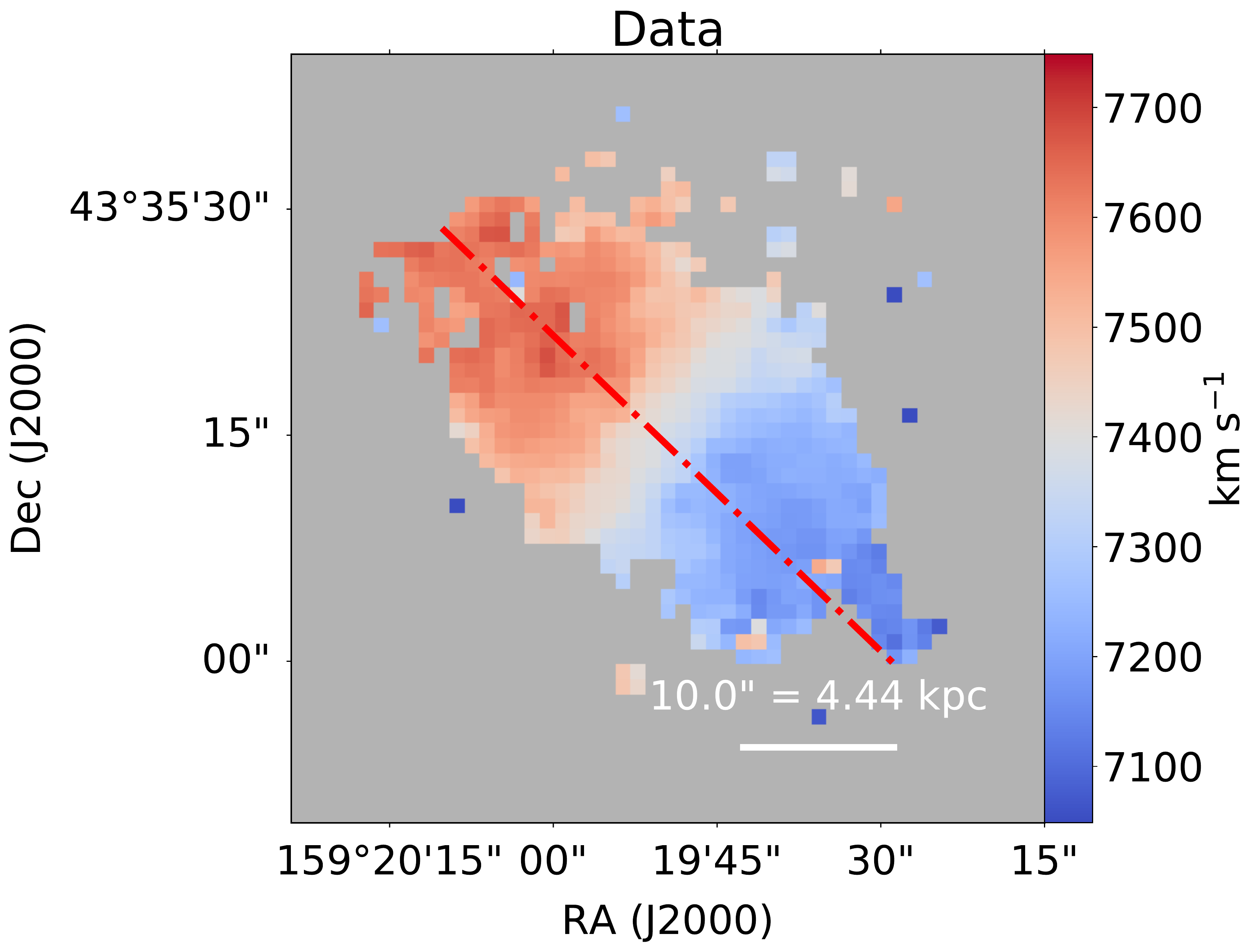}}
	\hfill
	\subcaptionbox{\label{fig: Halpha_vfit Gaussian}}{\includegraphics[width=0.49\linewidth]{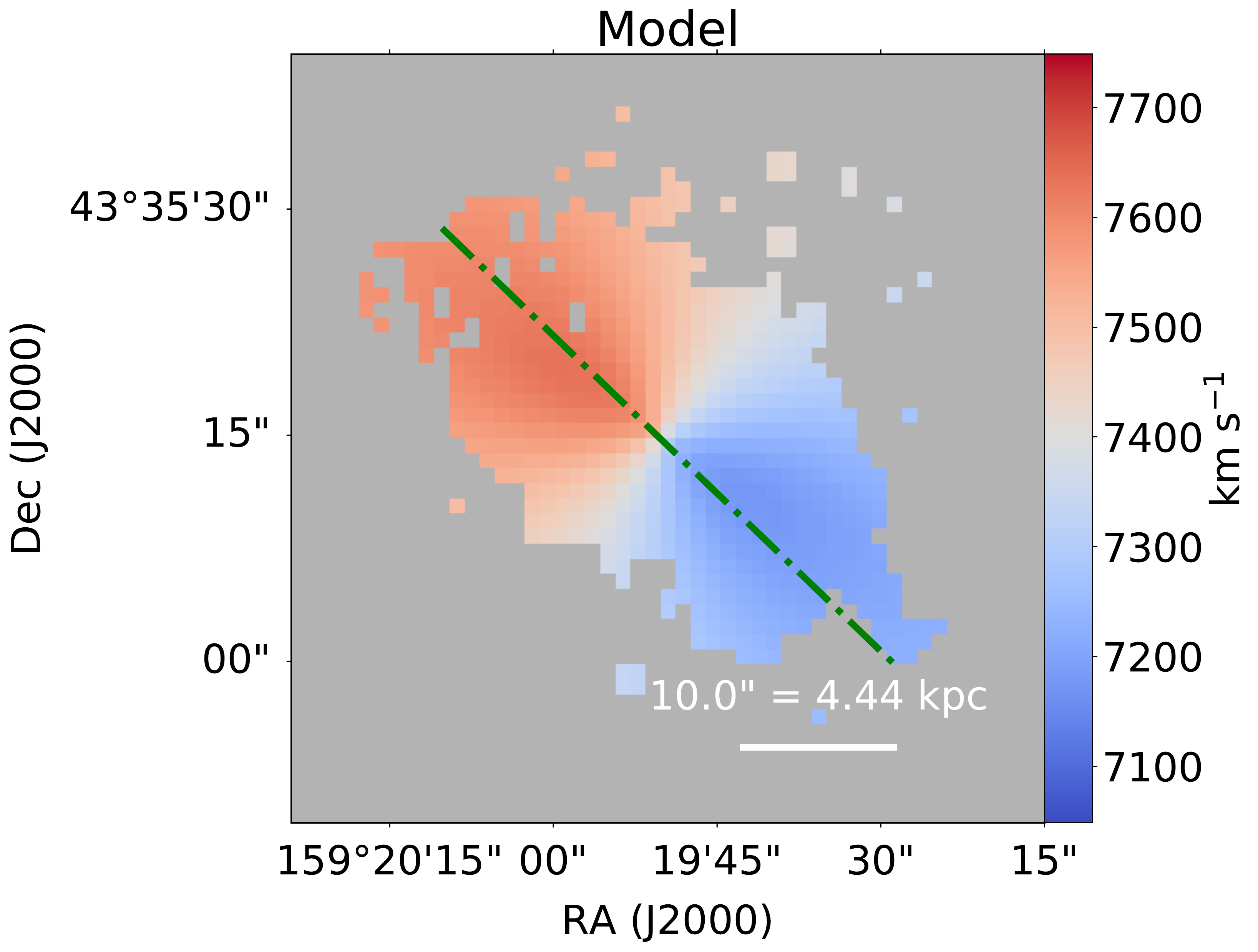}}
	\subcaptionbox{\label{fig: Halpha_vres Gaussian}}{\includegraphics[width=0.49\linewidth]{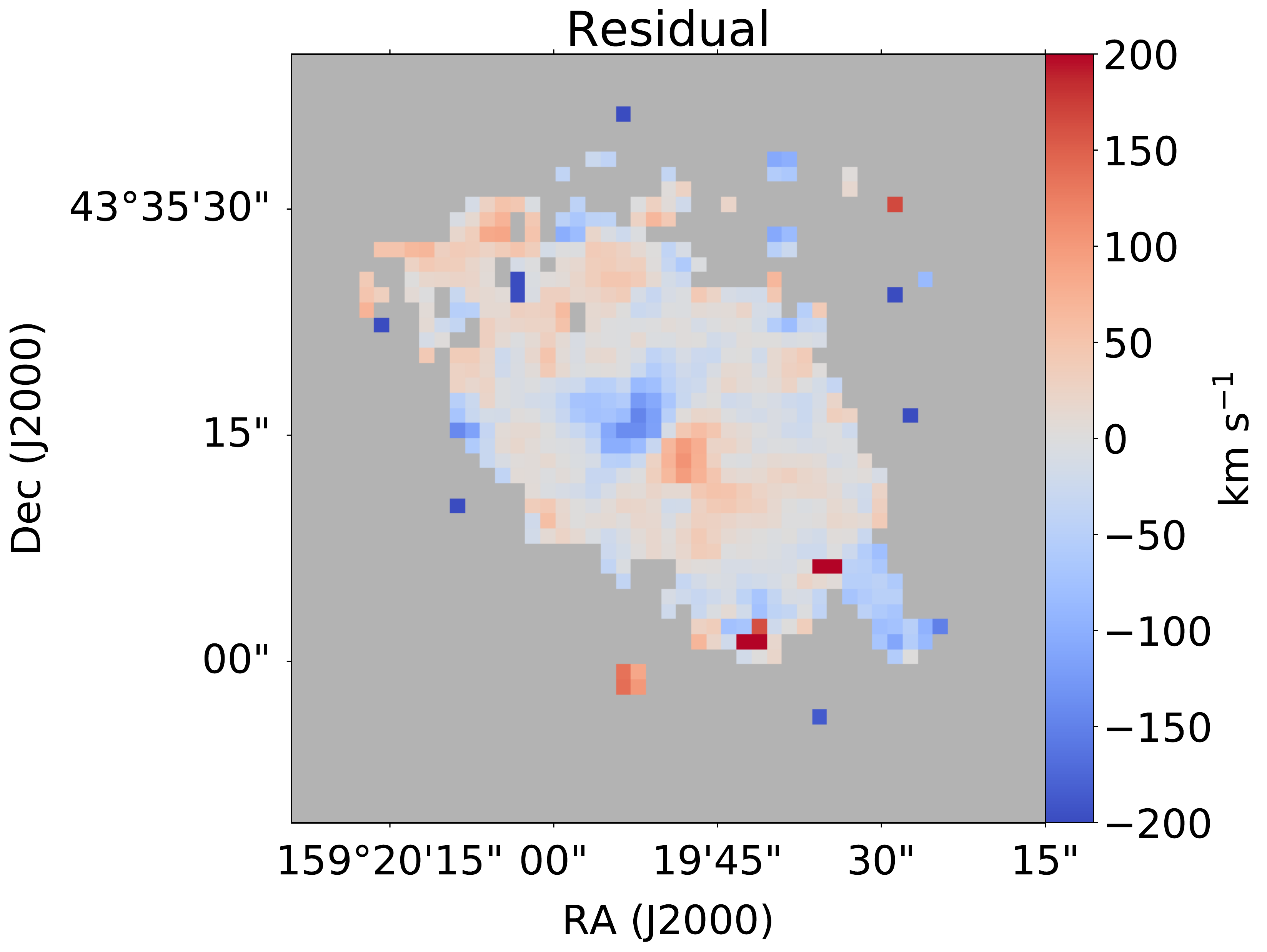}}
	\hfill
	\subcaptionbox{\label{fig: Halpha_xc Gaussian}}{\includegraphics[width=0.49\linewidth]{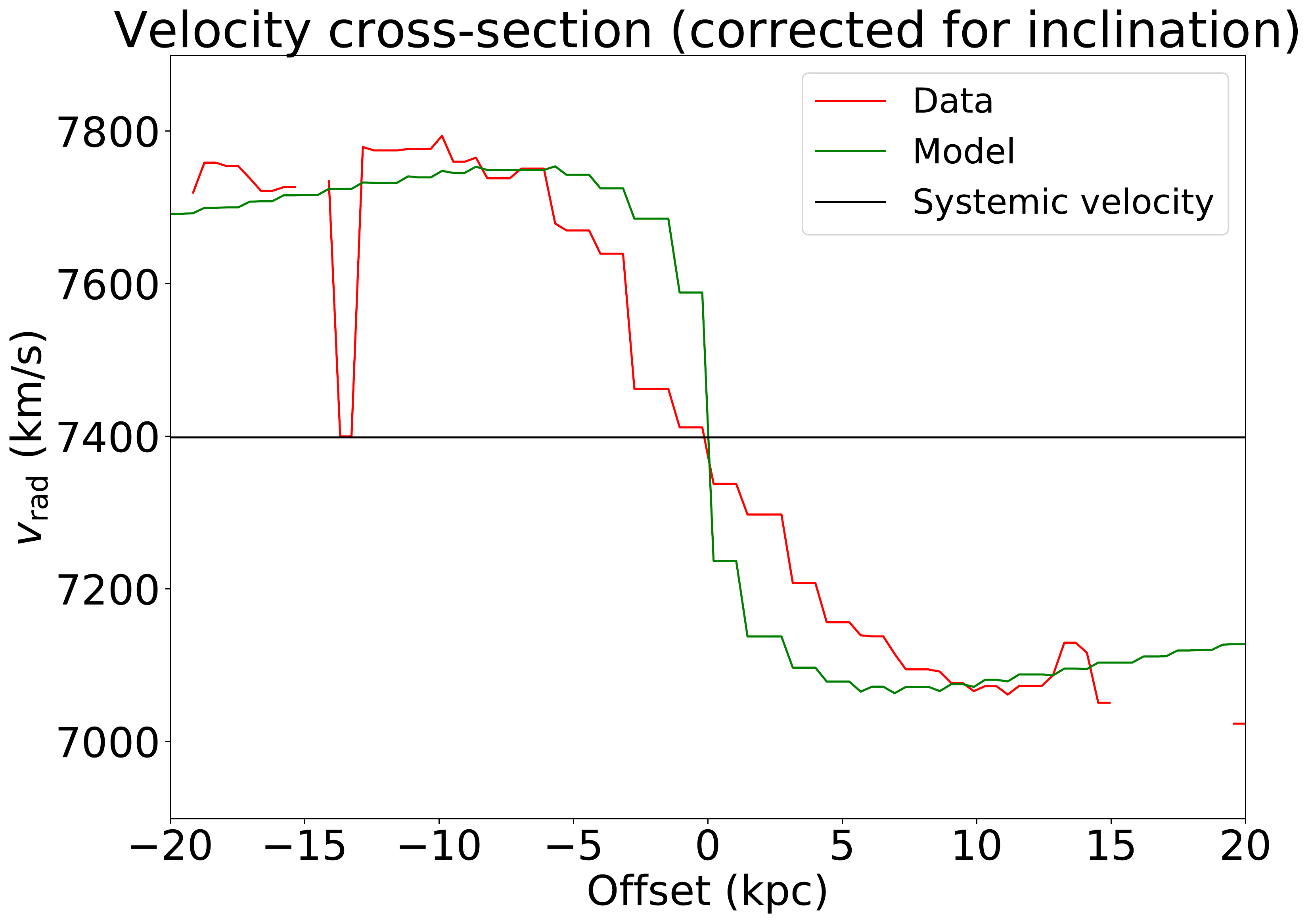}}
	\caption{(a) \ha{} radial velocity from the Gaussian fit, (b) S\'ersic disc model fit, (c) velocity residuals, and (d) a plot showing the radial velocity along the dashed green and red lines in Figs. (a) and (b), corrected for inclination. In Fig. (c), subtracting the model fit reveals a counter-rotating structure in the central region of the galaxy. The velocities in Figs. (a) and (b) are given with respect to the local standard of rest.}
	\label{fig: Halpha disc fit Gaussian}
\end{figure*}

Fig.~\ref{fig: Halpha disc fit Gaussian} shows our model fit and residuals to the radial velocity of the \ha{} disc. The $\chi^2 > 10$, indicating a poor fit for our rather simplistic disc model.
To ensure that any disturbed kinematics in the central region were not biasing the fit, we re-ran our fit with the central 2\,kpc masked. The best-fit parameters were almost identical in both cases, indicating that the kinematics in the nuclear region were not significantly biasing the fit.
The residual reveals a counter-rotating structure in the nucleus and a velocity gradient on larger scales in the disc. 
Although counter-rotating cores are not uncommon in ETGs~\citep[e.g.,][]{Bender1988}, this may also indicate that our simple disc model is a poor fit to the data. 

The \ha{} velocity dispersion (Fig.~\ref{fig: Halpha vdisp}) is clearly elevated within 2\,kpc of the nucleus, reaching values $\sigma \gtrsim 225\,\rm km\,s^{-1}$. 
Whilst the velocity dispersion in most parts of the disc does not exceed $100 \,\rm km\,s^{-1}$, the velocity dispersion within about 2\,kpc of the nucleus reaches values as high as $225 \,\rm km\,s^{-1}$.

To determine whether the enhanced central velocity dispersion could be due to beam smearing, we used our best-fit disc model to obtain line-of-sight (LOS) velocities, which we then used to create synthetic emission lines in each spaxel of a data cube. We modelled each emission line with an intrinsic Gaussian $\sigma$ of 75 km/s, to match the observed velocity dispersion in the outer regions of the gas disc, plus the instrumental resolution of PMAS. We spatially smoothed the cube to match the 2.5'' seeing of the CALIFA data, and then spatially binned the cube to match the spaxel size. Finally we added Gaussian noise to replicate the S/N in the CALIFA data. 
By measuring the velocity dispersion in each spaxel using our emission line fitting routine, we found that beam smearing was unable to reproduce the observed line widths in the central 2\,kpc. We hence rule out beam smearing as the sole cause of the elevated velocity dispersion in this region.

An alternative scenario to explain the high velocity dispersions and shock-like line ratios in the inner parts of the disk of UGC\,05771 could be accretion of gas from the surroundings of the galaxy. 
In principle, such an accretion event could also have triggered the nuclear activity. 
UGC\,05771 resides in a group environment with two massive galaxies within a few 100 kpc. Hence, a recent encounter or minor merger would not be implausible, although we do not see any evidence for that, e.g., in the stellar morphology of UGC\,05771.
Moreover, mass accretion rates required to fuel nuclear radio activity are very low, of order $10^{-3} \, \rm M_\odot \, yr^{-1}$ or less~\citep[e.g.,][]{Merloni2003}, so that a specific triggering event may not be necessary. 
We also do not see any signatures of infalling neutral gas, such as interstellar Na\,D absorption line components, and the rotation of the gas disk is overall very regular (Fig.~\ref{fig: Halpha vrad}); this is inconsistent with a recent merger event that would have been strong enough to stir up large parts of the gas in the nuclear regions.

An exception might be the small region within 2 kpc of the nucleus where we find non-circular motions, which could originate from a counter-rotating core, if this feature is not due to an interaction with extended jet plasma. 
However, for accreted gas to become bound to the disc, it cannot have a velocity much larger than the circular velocity of the disc, which is approximately $100 \,\rm km \, s^{-1}$ within the inner kpc (Fig.~\ref{fig: Halpha_xc Gaussian}). 
Basic energy conservation arguments would therefore make it unlikely that such gas can cause the observed velocity dispersions of $225 \,\rm km \, s^{-1}$, which are much greater, even when neglecting that much of the kinetic energy of the gas is being radiated away by line emission from shocked gas. 
We therefore do not consider accretion a strong alternative explanation, although detailed, high-resolution hydrodynamical simulations of rapid gas accretion events would be very useful to explore this alternative further.

\section{IRAM observations}\label{sec: IRAM data}

\subsection{Observations}\label{subsec: IRAM observations}

We obtained CO(1--0) and CO(2--1) spectra of UGC\,05771 with the IRAM 30\,m telescope through the Institut de Radio Astronomie Millim\'etrique (IRAM) Director's Discretionary Program D06-18 on February 25th, 2019. 
The source was observed with the Eight MIxer Receiver (EMIR), a wide-band heterodyne receiver, as part of the heterodyne pool. 
Both lines were observed together in dual-band mode with the FTS200 and the The Wideband Line Multiple Autocorrelator (WILMA) backends. 
They were placed in the UI sideband, with a tuning frequency corresponding to the expected observed frequencies $\nu_{\rm obs}$ of the lines at $z=0.02469$, $112.494\,\rm GHz$ and $224.983\,\rm GHz$ for CO(1--0) and CO(2--1) respectively. 
The IRAM beam FWHM is 21.8'' at 114\,GHz and 10.6'' at 228\,GHz.

We used Wobbler switching throws of 60'', which is larger than the size of CO line emission in UGC\,05771. The telescope was focused at the beginning of the observations on the bright quasar 1226$+$023, and then refocused every 3--4\,hr. Another quasar, 0923$+$392, which is near the source, was used to reposition the telescope on the sky every 2--3\,hr. Individual scans were 30\,s long, and we obtained a calibration after sequences of 12 scans, i.e., every 6\,minutes. The precipitable water vapour during observations varied between $1.5$ and $5\,\rm mm$. The total time on-source was 14580\,s.

\subsection{Data reduction}\label{subsec: IRAM data reduction}
 
The data were calibrated at the telescope, and reduced with the \textsc{CLASS} package of the IRAM \textsc{GILDAS} software~\citep{Gildas2013}. All scans were examined by eye, and first-order polynomials were used to correct the baselines, after masking the spectral range covered by the line. We then averaged the individual scans, combining both horizontal and vertical polarizations. We used the antennae efficiencies given on the EMIR website of $5.9 \,\rm Jy \, K^{-1}$ for the 3\,mm, and $7.5 \,\rm Jy \, K^{-1}$ for the 1.3\,mm data to translate the measured brightness temperatures into flux density units, i.e., Jy. The lines are detected with both backends. We used WILMA to extract the scientific results, because it has a spectral resolution better suited to measurements of extragalactic emission lines, and more stable baselines. The resulting root-mean-square (rms) noise is $5.3 \,\rm mJy$ and $7.5 \,\rm mJy$ for CO(1--0) and CO(2--1), respectively, for a spectral channel width of $5 \,\rm km \, s^{-1}$ at $112.5 \,\rm GHz$. 

\subsection{Results}\label{subsec: IRAM results}

\begin{figure}
	\centering
	\subcaptionbox{\label{fig: CO(1--0)}}{\includegraphics[width=1\linewidth]{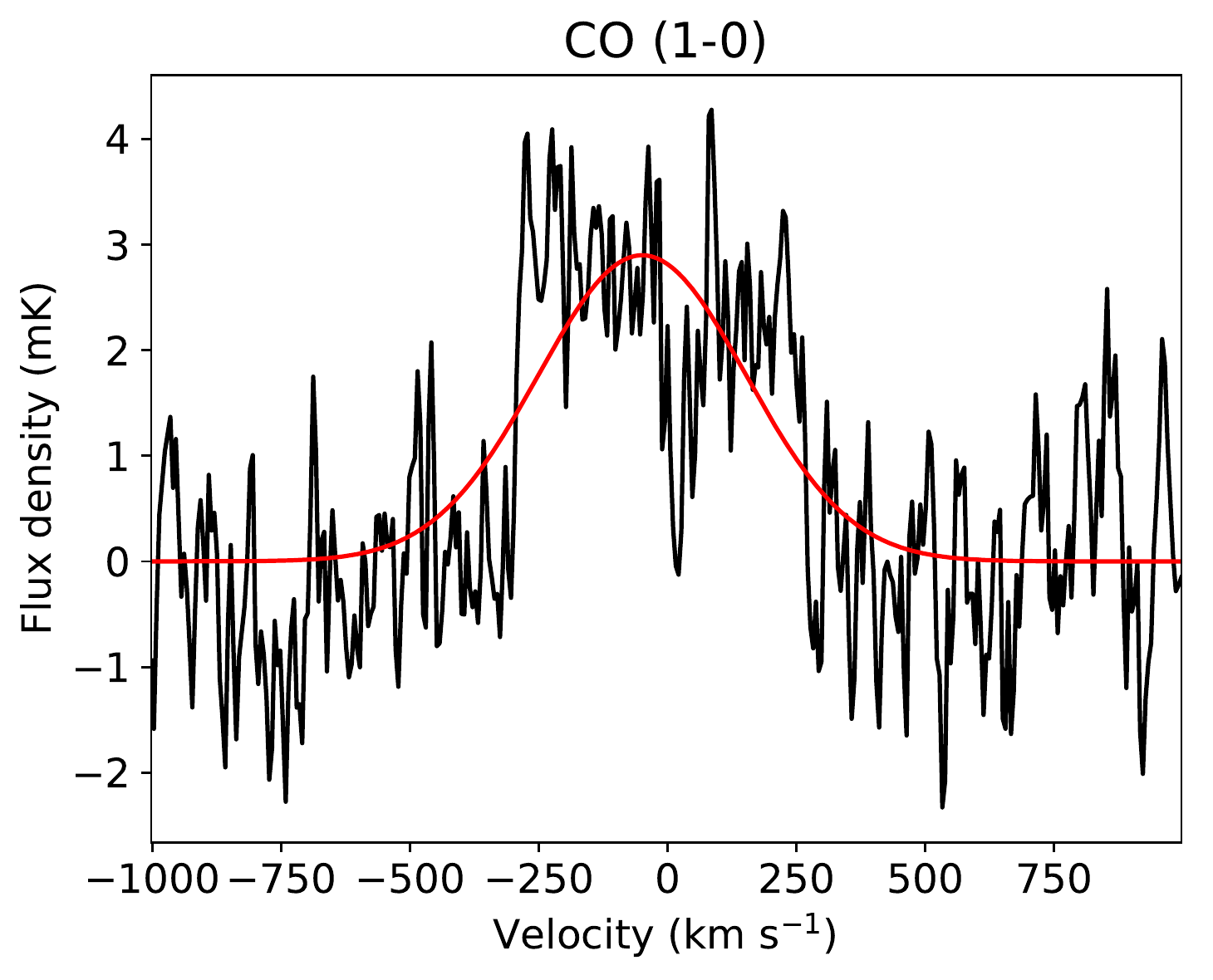}}
	\subcaptionbox{\label{fig: CO(2--1)}}{\includegraphics[width=1\linewidth]{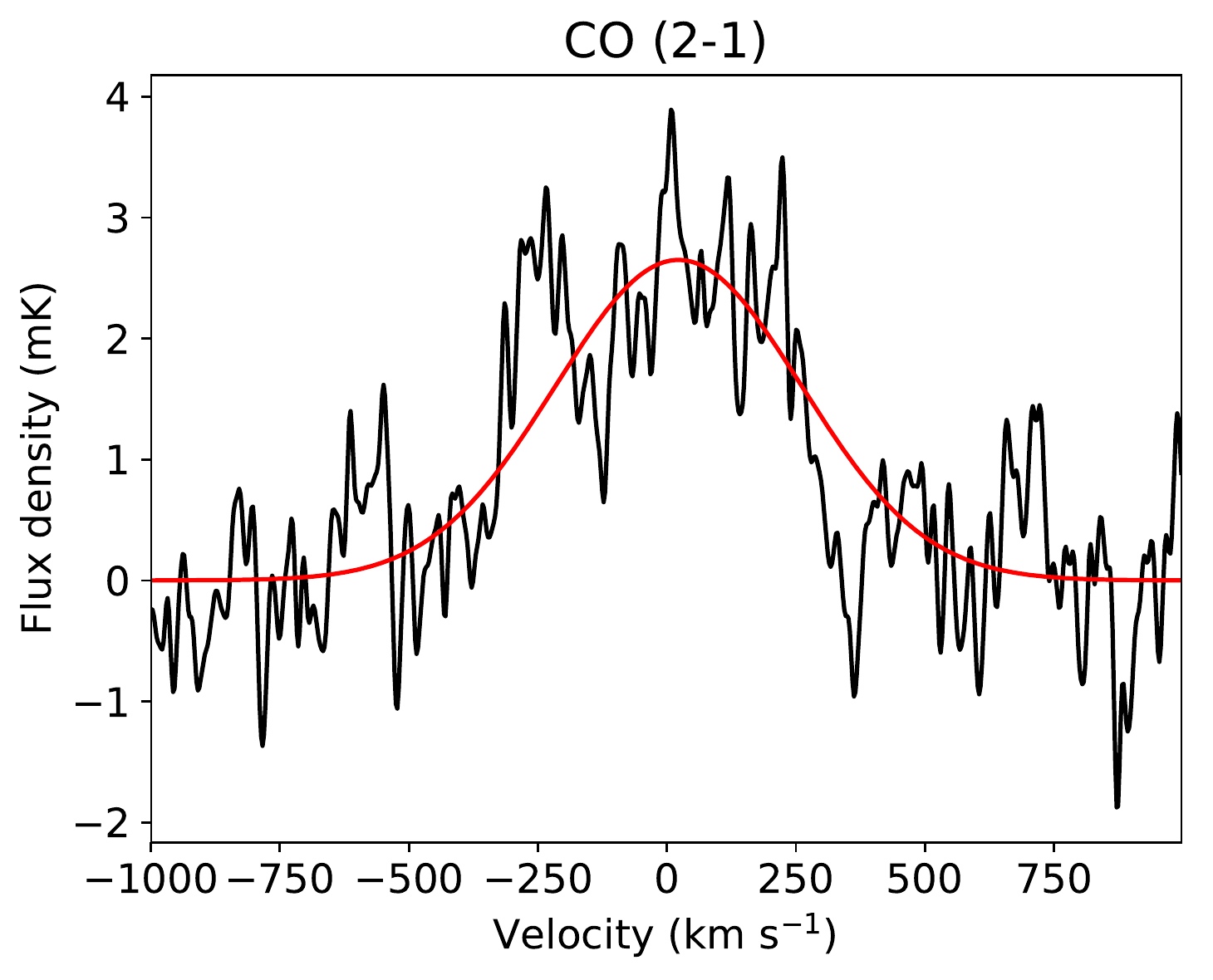}}
	\caption{(a) CO(1--0) and (b) CO(2--1) line profiles of UGC\,05771. Red lines show the best Gaussian fit to each line.}
	\label{fig: CO profiles}
\end{figure}

We show the CO(1--0) and CO(2--1) spectra of UGC\,05771 in Fig.~\ref{fig: CO profiles}, and the details of our observations and the fitted Gaussian profiles in Table~\ref{tab: CO data results}. Both lines were clearly detected at the expected redshifts.

The CO(1--0) line shows a possible double-peaked profile, with a total line width of $\rm FWHM = 476 \pm 23 \,\rm km \, s^{-1}$. The CO(2--1) line is somewhat broader, with $\rm FWHM = 562 \pm 26 \, \rm km \, s^{-1}$, and appears to have a more complex line profile, although this may be due to the lower S/N and more irregular baselines at 1.3\,mm. The CO(1--0) and CO(2--1) lines are centred around velocities of $49 \pm 13\,\rm km\,s^{-1}$ and $22\pm11 \,\rm km \, s^{-1}$ respectively, relative to the systemic redshift of UGC\,05771.
The broad widths of the CO line profiles are consistent with the radial velocities observed in the ionised gas disc ($v_{\rm rad} \approx 200 \,\rm km \, s^{-1}$, Fig.~\ref{fig: Halpha vrad}) suggesting the broad line widths are dominated by rotation, although additional broadening due to turbulence may be present.
The line fluxes are $8.6 \pm 0.4 \,\rm Jy \, km \, s^{-1}$ and $11.9 \pm 0.5 \,\rm Jy \, km \, s^{-1}$ for CO(1--0) and CO(2--1) respectively. 
The line ratio CO(2--1)/CO(1--0) $ = 2.5$ is consistent with those in the sample of powerful radio galaxies with CO observations by \citet{OcanaFlaquer2010}, although this ratio may be affected by the different beam sizes of the 30-m telescope at 3~mm and 1.3~mm, which would have an effect if the CO reservoir is more extended than the 10'' beam at 1.3~mm.

Although both the CO(1--0) and CO(2--1) line profiles are complex, we could not justify the use of a more complex model than a simple Gaussian profile due to the poor S/N.
To check the validity of this approach, we also estimated the integrated flux by summing the spectra between $\pm 500 \, \rm km\,s^{-1}$; for both the CO(1--0) and (2--1) lines, this method produced values within the $1\sigma$ errors of the fluxes from the Gaussian fit.

\begin{table*}
	\caption{Observational results for CO: Rest-frame and observed
		frequency, velocity offset to $z=0.02469$, FWHM line width, peak
		brightness temperature, and integrated line flux.}
	\label{tab: CO data results}
	\begin{tabular}{lcccccc}
		\hline
		\textbf{Line} & $\nu_0$ (GHz) & $\nu_{\rm obs}$ (GHz) & $\Delta v$ (km s$^{-1}$) & \textbf{FWHM} (km s$^{-1}$) & $T_{\rm pk}$ (mK) & \textbf{Flux $I_{\rm CO}$} (Jy km s$^{-1}$) \\
		\hline
		CO(1--0) & $115.271$ & $112.4755 \pm 0.0013$ & $-49.4 \pm 4.7$ & $476 \pm 23$ & $2.9 \pm 0.1$ & $8.6 \pm 0.4$ \\
		CO(2--1) & $230.538$ & $224.9662 \pm 0.0009$ & $22.4 \pm 11.5$ & $562 \pm 26$ & $2.65 \pm 0.1$ & $11.9 \pm 0.5$ \\
		\hline  
	\end{tabular}
\end{table*}

To estimate the molecular gas mass from the integrated CO(1--0) line flux $I_{\rm CO}$, we first estimated the total CO line luminosity $L_{\rm CO}^\prime$ using eqn. 3 of \citet{Solomon1997},
\begin{equation}
L_{\rm CO}^\prime = 3.25 \times 10^7 I_{\rm CO} \left[\nu_0 (1+z)\right]^{-2} D_L^2 (1+z)^{-3}
\end{equation}
where $L_{\rm CO}^\prime$ has units K km s$^{-1}$\,pc$^2$, the luminosity distance $D_L$ is given in Mpc, the rest-frame emission line frequency $\nu_0$ is given in GHz and $I_{\rm CO}$ is given in Jy km s$^{-1}$. 
As is standard for nearby ETGs, we used a Milky Way conversion factor between CO luminosity and \hh{} gas mass, corresponding to $\alpha_{\rm CO} = 4.6 \rm \, M_{\odot}$ / [K km s$^{-1}$\,pc$^2$], which gives a molecular gas mass $M_{\rm H_2, \, \rm CO} = 1.1 \pm 0.4 \times\ 10^9 \,\rm M_{\odot}$.

\subsubsection{Surface density of dense gas}\label{subsubsec: IRAM gas surface density}

Resolved CO observations suggest that CO emission traces dust in ETGs~\citep{Alatalo2013}.
Hence, because our CO observations are not spatially resolved, we estimated the dense gas mass surface density $\Sigma_{\rm gas}$ using the extinction as a proxy for the distribution of dense gas. 
At the frequency of the observed CO(1--0) transition, the FWHM of the IRAM beam is $\rm FWHM = \left(2460/\nu_{\rm obs}[\rm GHz]\right)\text{''} = 21.9\text{''}$, slightly smaller than the size of the ionised gas disc.
As shown in Fig.~\ref{fig: A_V map}, the IRAM beam covers roughly the same area over which we are able to measure $A_V$; over this region, $A_V$ varies from $0 - 2$ magnitudes and has a clumpy distribution, showing no clear trends with radius.
Hence it is reasonable for us to assume that the CO emission is uniformly distributed across the IRAM beam, which gives $\Sigma_{\rm gas} = 15 \pm 5 \,\rm M_\odot \,\,pc^{-2}$, consistent with our $A_V$-based estimate ($\Sigma_{\rm gas} = 10.0 \pm 0.5 \, \rm M_\odot \,\,pc^{-2}$.). 
We use this gas mass surface density estimate in Section~\ref{subsec: SFR discussion} to investigate whether the jets are inducing negative feedback in UGC\,05771.

\section{Discussion}\label{sec: Discussion}

In this section, we present our interpretation of our OSIRIS, CALIFA and IRAM results. 
First, we argue that the shocked near-IR and optical line emission is due to jet-ISM interactions, and therefore that the radio source is much more extended than its apparent size in existing VLBI observations.
We then use our CO observations and SFR estimates to determine whether the extended jet plasma is inhibiting star formation in UGC\,05771.
Finally, we constrain the properties of the density distribution of the ISM in UGC\,05771 and estimate the age of the radio source.

\subsection{Evidence for jet-ISM interaction in UGC\,05771}
\label{subsec: What is causing the [Fe II] and H2 emission?}

Using our OSIRIS data (Section~\ref{sec: OSIRIS data}), we detected shocked molecular and ionised gas at radii of $\approx 200\,\rm\,pc$, whereas the CALIFA data (Section~\ref{sec: CALIFA data}) revealed shocked ionised gas at radii of $\approx 1\,\rm\,kpc$. Here, we argue that both the optical and near-IR line emission is caused by the radio jets.

In Section~\ref{sec: [Fe II] emission}, we showed that SNe explosions are unable to reproduce the observed \feii{} line luminosity. 
Meanwhile, the \feii{}, \hh{} and \ha{} luminosities represent a few per cent of the estimated jet power ($4.2 \times 10^{41}\,\rm erg\,s^{-1}$), meaning it is energetically plausible for the jets to power the line emission via radiative shocks.
The \feii{} exhibits a sharp velocity gradient of $\approx 200 \,\rm km\,s^{-1}$ and has broad line widths of up to $230\,\rm km\,s^{-1}$ (Fig.~\ref{fig: [Fe II] maps}), indicating this gas is being excited by an energetic process. 
The radial velocities expected due to galactic rotation at these radii are insufficient to explain the velocity gradient in the \feii{}. 
Meanwhile, there is no coherent rotation in the warm \hh{}, with most of the line emission being blueshifted (Fig.~\ref{fig: H2 maps}), suggesting this material is being ejected from the nucleus.
The radial velocities of both the \feii{} and the \hh{} are too small for the gas to escape the host galaxy's potential, with $v_{\textrm{esc}} \approx 900\,\rm km\,s^{-1}$ at comparable radii. 
We therefore conclude that both the \hh{} and the \feii{} emission probe gas being accelerated out of the nucleus by the jets, entrained in a `stalling wind' that will not escape the host galaxy.
Similar phenomena have previously been observed in NGC\,1266~\citep[e.g.,][]{Alatalo2015} and in 3C\,326\,N~\citep{Nesvadba2010,Nesvadba2011}, in which only a fraction of the emission-line molecular gas accelerated by the jets exceeds the escape velocity of the host galaxy.

In our CALIFA data we detected a kpc-scale region of shocked gas with an elevated velocity dispersion in the vicinity of the nucleus. 
We have shown that beam smearing is unable to reproduce the observed velocity dispersion (Section~\ref{subsec: CALIFA emission line analysis}), and that the line ratios in this region are consistent with shocks (Fig.~\ref{fig: BPT_vdisp}). 
We also demonstrated that it is highly unlikely that UGC\,05771 is accreting enough material to drive these shocks, given the relatively ordered kinematics in the ionised gas disc, and the lack of any signatures of infalling gas (Section~\ref{subsubsec: CALIFA ionised gas kinematics}).
Hence, we conclude that\,kpc-scale jet plasma must be responsible for this line emission.

However, the apparent size of the radio source is $9\,\rm\,pc$ (e.g., Fig.~\ref{fig: VLBI: 1.665 GHz}). This is inconsistent with our hypotheses that the jets power the \hh{} and \feii{} emission, which extend $\approx 200 \,\rm\,pc$ from the nucleus, and the \ha{} emission, which extends $\approx 1 \,\rm\,kpc$ from the nucleus. We address this in the following section.

\subsection{How extended is the radio source?}

\begin{figure}
	\centering
	\includegraphics[width=1\linewidth]{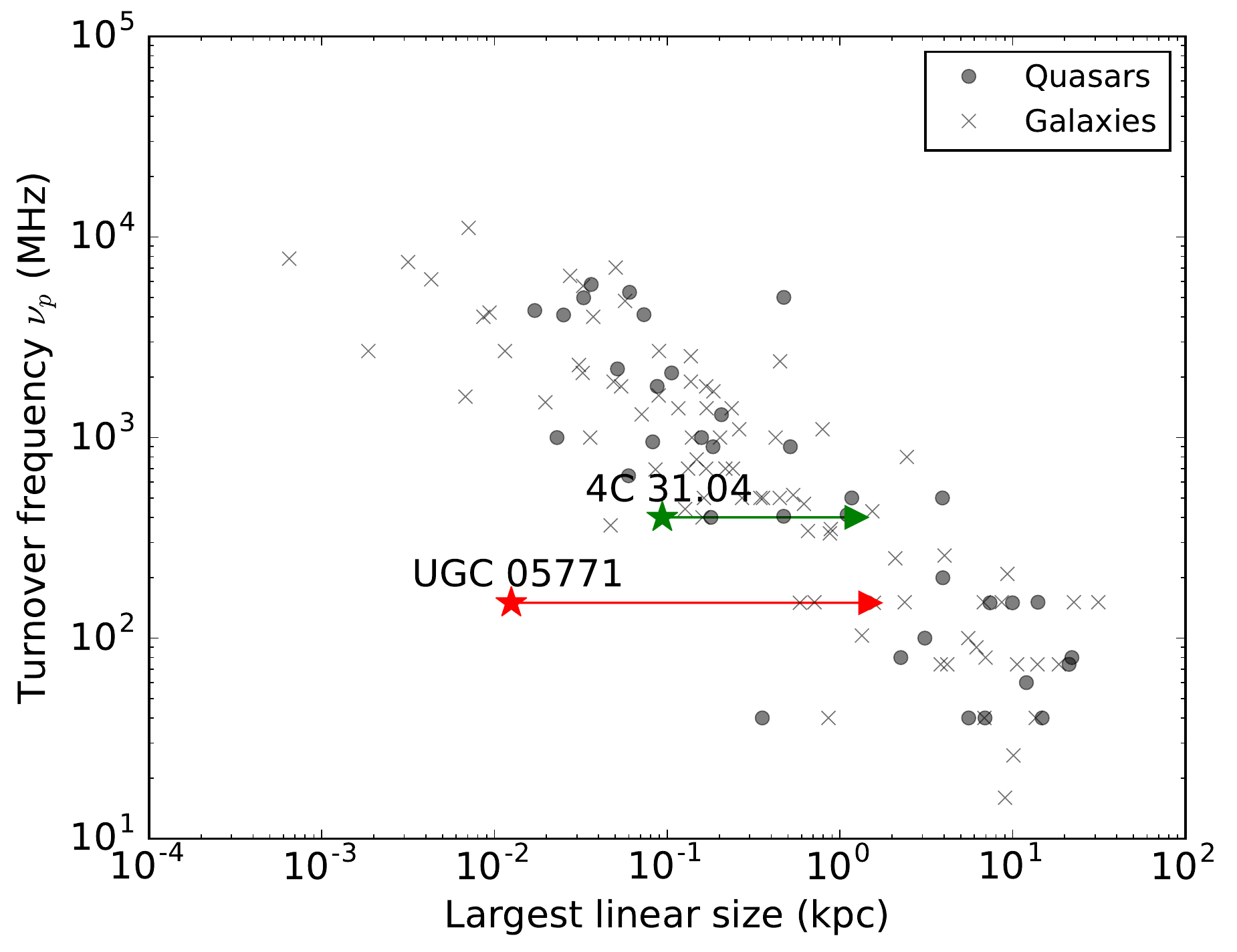}
	\caption{The peak frequency versus largest linear size for the catalogue of GPS and CSS sources compiled by \citet{Jeyakumar2016}, illustrating the strong anticorrelation between turnover frequency and linear size.
		The stars indicate the linear sizes, derived from VLBI observations, of 4C 31.04~\citep[][green]{Zovaro2019} and UGC\,05771 (red) presented by \citet{Giroletti2003} and \citet{deVries2009} respectively. Meanwhile, the arrows indicate the approximate extent of ro-vibrational \hh{} emission in in 4C 31.04 and optical emission line gas with line ratios consistent with shocks (Fig.~\ref{fig: BPT_vdisp}), which may trace extended, low surface brightness jet plasma that is resolved out in the VLBI images.}
	\label{fig: linear size vs. peak frequency plot}
\end{figure}

The turnover frequency of GPS and CSS sources is strongly anticorrelated with the radio size of the source~\citep{O'Dea&Baum1997}.
In Fig.~\ref{fig: linear size vs. peak frequency plot}, we show this correlation for the sample of GPS and CSS sources compiled by \citet{Jeyakumar2016}. We also indicate UGC\,05771 on the plot (red star), using the linear size from the VLBI observations~\citep[][Figs.~\ref{fig: VLBI: 1.665 GHz} and \ref{fig: VLBI: 4.993 GHz}]{deVries2009}.
UGC\,05771 is offset from the correlation by nearly two orders of magnitude, indicating it is possible that the radio source is in fact $\gtrsim 100$ times more extended than its size in existing VLBI observations.

If the jets are oriented at a small angle $\theta$ to the line of sight, the source will appear much more compact than it truly is.
Such beamed sources generally exhibit high flux variability. However, \citet{Snellen2004} find that UGC\,05771 does not have significant flux variability at 5 and 8.4\,GHz. We therefore conclude that the compactness of the radio source is not an orientation effect.
Why, then, does it appear to be so small? 

The linear size of the radio source associated with UGC\,05771 is based on the VLBI observations of \citet{deVries2009} at 1.665\,GHz.
Referring to Fig.~\ref{fig: radio spectrum}, these VLBI observations recover only $20$ to $30\,\rm per\,cent$ of the unresolved single-dish fluxes at comparable frequencies. This implies that $70 - 80\,\rm per \, cent$ of the radio emission at these frequencies is emitted on scales larger than the spatial cutoff frequency of the VLBI observations.

The presence of bright, compact radio structures embedded in extended regions of low surface brightness radio emission is a key signature of jets in the ``flood-and-channel phase'' of evolution~\citep{Sutherland&Bicknell2007}.
In this phase, emergent jets split into multiple streams as they interact with a clumpy medium. Whilst the weaker streams percolate isotropically and form a bubble that drives a shock into the surrounding medium, the main jet stream emits brightly in the radio as it interacts with the ISM impeding its passage.
Synthetic radio surface brightness images of these simulations~\citep[e.g., fig. 9 of ][]{Zovaro2019} show that the bubble can expand rapidly, enabling low-surface brightness plasma to be much more extended than the brightest radio structures. 
The low flux completeness of the\,pc-scale VLBI observations, coupled with the\,kpc-scale shocked gas, therefore suggests the jets in UGC\,05771 are in the ``flood-and-channel phase'', which we have previously observed in the CSS source 4C\,31.04~\citep{Zovaro2019}.
The possible presence of a counter-rotating core (Fig.~\ref{fig: Halpha_vres Gaussian}) may additionally be a signature of jets interacting with a clumpy disc~\citep{Mukherjee2018b}, a phenomenon also observed in IC~5063~\citep{Morganti2015,Mukherjee2018a}.

\subsection{Are the jets inducing negative feedback in UGC\,05771?}
\label{subsec: SFR discussion}

Even if the jets in UGC\,05771 are not powerful enough to prevent star formation by ejecting gas from the host galaxy's potential, our optical and near-IR observations indicate that there is kpc-scale jet plasma driving shocks and turbulence into the ISM. In this section, we use the Kennicutt-Schmidt (KS) relation~\citep{Kennicutt1998} to determine whether the galaxy has a noticeable offset from that relation that could be attributed to the interactions with the radio jet. 

Fig. 15 shows where UGC\,05771 lies with respect to the KS relation (assuming a Salpeter IMF). To estimate $\Sigma_{\rm gas}$ in UGC\,05771, we use our CO observations; for the SFR surface density $\Sigma_{\rm SFR}$, we show our estimates computed using both the total \ha{} flux (empty red triangle) and the \ha{} flux only from spaxels with emission line ratios lying in the intermediate and star-forming regions of the optical diagnostic diagrams (filled red triangle). We favour the latter value, as the first is contaminated by gas where shocks, AGN photoionisation and/or photoionisation from post-AGB stars dominates, as indicated by the line ratios.

For comparison, we also show galaxies from two different samples: 15 early-type galaxies from \citet{Martig2013}, which do not host AGN, but may show signatures of morphological quenching, and 16 bright, gas-rich radio galaxies ('molecular hydrogen emission galaxies' (MOHEGs)) from \citet{Lanz2016}, which contain warm, shocked molecular gas, presumably from interactions with the radio jet~\citep{Ogle2010,Nesvadba2010}. The star-formation rates in these samples were derived from PAHs and the far-infrared dust emission, respectively, which may imply small systematic uncertainties of-order $0.1-0.2$ dex compared to UGC\,05771. 

We also show the original sample of spiral and starburst galaxies from \citet{Kennicutt1998}, where the SFRs have been estimated using \ha{} and FIR fluxes respectively, both adjusted for a Salpeter IMF. Adopting the $\rm SFR(\rm H\upalpha)$ estimate only using the intermediate and star-forming spaxels, UGC\,05771 is shifted by a factor of 9 from the KS relationship of ordinary star-forming galaxies, corresponding to a shift of about 1 sigma, and in the regime covered by the samples of \citet{Martig2013} and \citet{Lanz2016}. However, as discussed in Section~\ref{subsubsec: Halpha SFR estimate}, our $\Sigma_{\rm SFR}$ estimate should be interpreted as an upper limit due to contamination from shocks, AGN photoionisation and/or post-AGB stars. It is therefore feasible that UGC\,05771 is even more strongly offset. Spatially resolved observations would be needed to see whether this offset of the global, source averaged gas and star-formation surface densities heralds a more pronounced, local decrease in star formation efficiency, or whether a slight global decrease is happening in UGC\,05771. This would also help to distinguish whether such a decrease would be due to interactions with the radio jet, or other quenching mechanisms, e.g., morphological quenching. Regardless of the detailed mechanism in place, however, our results do show that UGC\,05771 has a slightly lower star formation rate than would be expected from the Kennicutt-Schmidt law at face value. 

\begin{figure}
	\centering
	\includegraphics[width=1\linewidth]{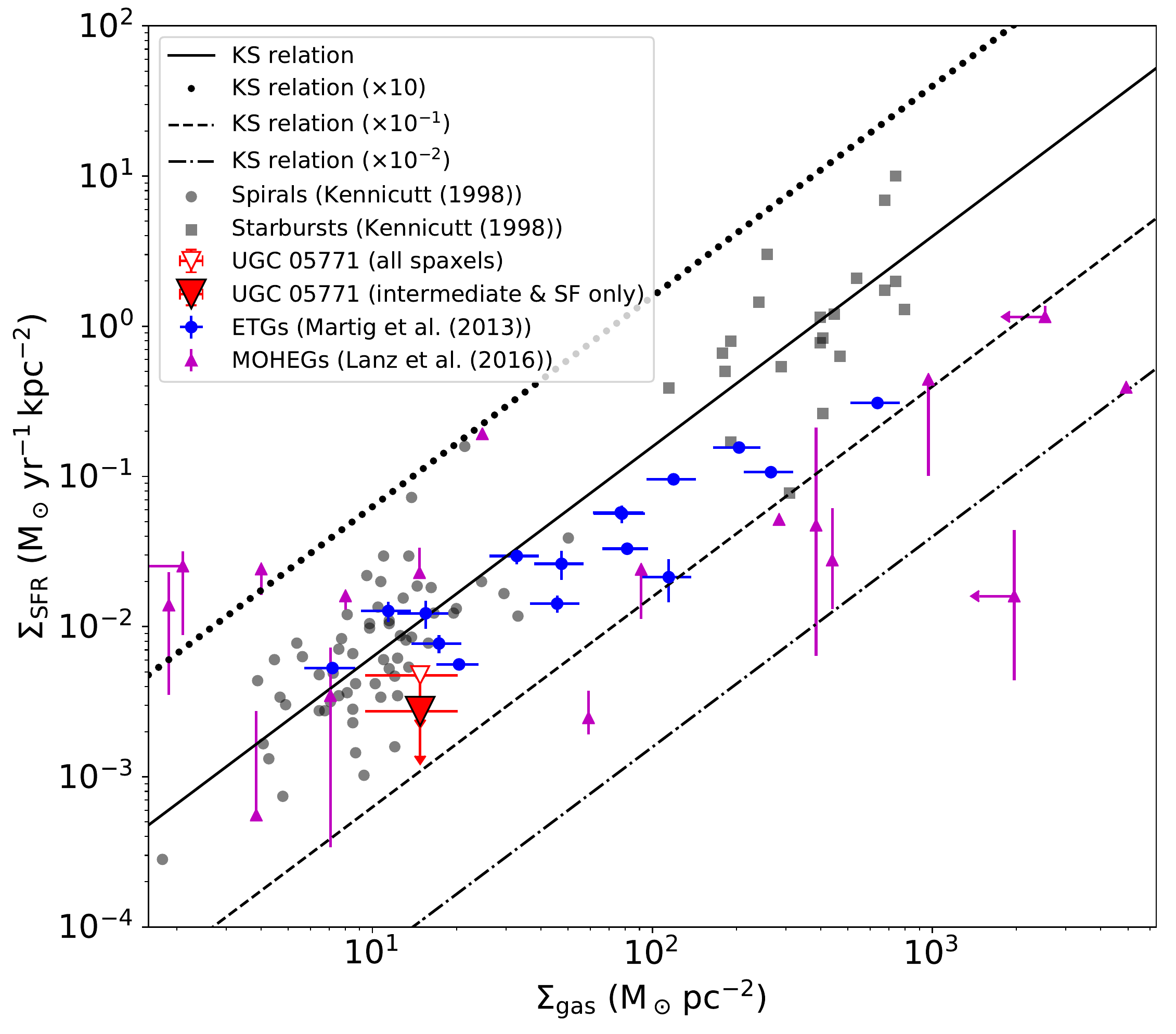}
	\caption{
			The Kennicutt-Schmidt (KS) relation.
			The black line represents the KS relation for a Salpeter IMF. We also show lines which display star forming efficiencies of $10\times$ (dotted line), $0.1\times$ (dashed line) and $0.01\times$ (dot-dashed line) that of the KS relation.
			We indicate UGC\,05771 with $1\sigma$ error bars for $\Sigma_{\rm gas}$ and $\Sigma_{\rm SFR}$, where we have adopted the \ha{}-based SFR surface density computed using all spaxels within the IRAM beam (empty red triangle) and computed using only those spaxels with line ratios that lie in the intermediate or star-forming regions of the ODD shown in Fig.~\ref{fig: ODDs} (filled red triangle). The blue points are the sample of quiescent ETGs of \citet{Martig2013}, where the SFRs have been estimated using the $8 \,\rm \upmu m$ flux, and the magenta points are the sample of radio galaxies of \citet{Lanz2016}.
			The grey circles and squares are the sample of spiral and starburst galaxies of \citet{Kennicutt1998}.
			The SFRs for all samples have been adjusted for a Salpeter IMF.
		}
	\label{fig: KS relation}
\end{figure}

\subsection{Constraining the parameters of the ISM density distribution}\label{subsec: Constraining the parameters of the ISM density distribution}

	Free-free absorption (FFA) of synchrotron radiation by an inhomogeneous ionised medium enshrouding the jet plasma can replicate the characteristic spectrum of GPS and CSS sources~\citep{Bicknell2018}. Under this paradigm, the frequency of the spectral peak is sensitive to the density probability distribution function (PDF) of the ISM. Here, we use the method of \citet{Zovaro2019} to infer the properties of the density PDF in UGC\,05771; for full details of the method, we refer the reader to their section 5.4.

	We modelled the ISM as having a log-normal density distribution, which is appropriate for turbulent media~\citep{Nordlund&Padoan1999,Federrath&Klessen2012}, and assumed the absorbing medium is a slab with depth $L = 2\,\rm\,kpc$, corresponding to the region of elevated velocity dispersion in the ionised gas (Fig.~\ref{fig: Halpha vdisp}). 
	Eqn. 17 of \citet{Zovaro2019} gives the the free-free optical depth $\tau_\nu$ as a function of ISM parameters shown in Table~\ref{tab: parameters used in ISM calculation} and the expected value of $n^2$ in the slab $E(n^2)$. We estimated $E(n^2)$ by setting $\tau_\nu = 1$ at the spectral peak $\nu_p$, and used eqns. 11--13 and 18 of \citet{Zovaro2019} to estimate the mean density $\bar{n}$ and the standard deviation $\sigma$ of the density PDF, shown in in Table~\ref{tab: parameters used in ISM calculation}. The low mean density ($\sim 1\,\rm cm^{-3}$) is consistent with the low-frequency spectral turnover of UGC\,05771, whereas the inferred high turbulent Mach number and the assumption of a log-normal distribution gives a large value for $E(n^2)$. 

\begin{table}
	\caption{Parameters used in determining the parameters of the log-normal density distribution. Output parameters are denoted with daggers ($^\dagger$). Fractional abundances were estimated using a \textsc{Mappings V}~\citep{Sutherland2013} model grid with non-equilibrium cooling and solar abundances.}
	\begin{tabular}{c c c}
		\hline
		\textbf{Parameter} & \textbf{Symbol} & \textbf{Value} \\
		\hline
		Peak frequency & $\nu_p$ & $150 \,\rm MHz$ \\
		Depth of absorbing slab & $L$ & $2\,\rm\,kpc$ \\
		Temperature & $T$ & $10^4\,\rm K$ \\
		Mean molecular mass & $\mu$ & 0.66504 \\ 
		Electron fractional abundance & $n_e / n$ & $0.47175$ \\
		H$^+$ fractional abundance & $n_{\rm H^+} / n$ & $0.41932$ \\
		He$^+$ fractional abundance & $n_{\rm He^+} / n$ & $0.024458$ \\
		He$^{++}$ fractional abundance & $n_{\rm He^{++}} / n$ & $0.013770$ \\
		Line-of-sight velocity dispersion & $\sigma_g$ & $225\,\rm km\,s^{-1}$ \\
		Sound speed & $c_s$ & $11.21 \,\rm km\,s^{-1}$ \\
		Turbulent Mach number & $\mathcal{M}$ & $34.75\,\rm km\,s^{-1}$\\
		Turbulent forcing parameter & $b$ & $0.4$ \\
		Ratio of thermal to magnetic pressure & $\beta$ & $1$ (equipartition) \\
		\hline
		Expected value of $n^2$$^\dagger$ & $E(n^2)$ & $116.7\,\rm cm^{-6}$\\
		Mean density$^\dagger$ & $\bar{n}$ & $1.094\,\rm cm^{-3}$ \\
		Density variance$^\dagger$ & $\sigma^2$ & $115.5\,\rm cm^{-6}$ \\
		\hline
	\end{tabular}
	\label{tab: parameters used in ISM calculation}
\end{table} 

We then estimated the age of the radio source using the extent of the shocked gas, our new density estimate and the jet power. 
Assuming that the jet-driven bubble evolves adiabatically, the time taken $t_b$ for the bubble to reach its current size $R_b$ if inflated by a jet with power $L_{\rm jet}$ expanding into a uniform medium with density $\rho$ is given by~\citep{Bicknell&Begelman1996} 
\begin{equation}
t_b = \left( \frac{ 384 \upi }{ 125 } \right)^{1/3} \rho^{1/3} L_{\rm jet}^{-1/3} R_b^{5/3}
\end{equation}
where we adopted $R_b = 2\,\rm\,kpc$ and a density $n = 1\,\rm cm^{-3}$ consistent with our FFA model.
This gives an age of approximately $19 \,\rm Myr$.

\section{Conclusion}\label{sec: Conclusion}
We have studied jet-ISM interactions in the low-power CSS source UGC\,05771 in order to search for evidence that the jets are inhibiting star formation in the host galaxy.

We analysed the sub-kpc scale circumnuclear gas of the host galaxy to search for signatures of jet-ISM interactions using near-infrared integral field spectroscopy from OSIRIS. 

We detected ro-vibrational \hh{} emission in the central $200\,\rm\,pc$ of the host galaxy that probes shock-heated molecular gas at $T \approx 5000\,\rm K$. 
We also detected \feiilong{} emission in the same region with luminosity too high to be explained by SNe explosions, whereas the jet power is sufficient to power both the \feii{} and \hh{} luminosities, leaving shocks induced by the jets as the most likely cause. 
The kinematics of both the \hh{} and \feii{} emission lines imply that the jets are accelerating material out of the nucleus to velocities insufficient to expel the gas from the host galaxy potential, creating a `stalling wind'.

We analysed the properties of the\,kpc-scale optical emission line gas in the galaxy using optical integral field spectroscopy from the CALIFA survey. The host galaxy has a disc of ionised gas approximately 20\,kpc in diameter. 
Line ratios, broad line widths and disturbed kinematics in the innermost 2\,kpc cannot be explained by beam smearing or by accretion, indicating that the gas in this region is being shocked and disturbed by the jets.
Further observations with higher spectral resolution are required to confirm the presence of multiple kinematic components in the emission lines that would arise from jet-ISM interactions.

Both the CALIFA and OSIRIS data show that the jets are interacting strongly with the ISM out to\,kpc radii, in apparent contradiction to the\,pc-scale structure revealed by VLBI imaging. We proposed that UGC\,05771 in fact hosts a\,kpc-scale radio source, and that the extended jet plasma is either resolved out by these observations or has too low a surface brightness to be detected, consistent with the location of UGC\,05771 on the peak frequency-size correlation.
	
To determine whether the jets are inhibiting star formation in UGC\,05771, we obtained IRAM observations of CO(1--0) and CO(2--1), and found $M_{\rm gas} = 1.1 \pm 0.4 \times 10^{9} \,\rm M_{\odot}$ and a mean surface density $\Sigma_{\rm gas} = 15 \pm 5 \,\rm M_{\odot} \,\,pc^{-2}$. 
Although we found that it is possible that UGC\,05771 is significantly offset from the KS relation, we were unable to confirm whether negative feedback is taking place due to systematic uncertainties in our \ha{}-based SFR estimate due to contamination by shocks and evolved stars.

The fact that we have observed signatures of jet-ISM interactions out to\,kpc radii in both 4C~31.04 and UGC\,05771 suggests that diffuse, low surface brightness radio plasma that is not visible in VLBI observations may be common in compact radio galaxies. 
Our observations have shown that this radio plasma interacts strongly with the ISM, heating and injecting turbulence, and potentially inhibiting star formation. 
This finding demonstrates that young radio sources with seemingly compact jets may have a substantial impact on the star formation of their host galaxy, having important implications for the role that jets play in galaxy evolution.

\section*{Acknowledgements}
We would like to thank the anonymous reviewer for their insightful comments, and Anne Medling, Alec Thomson and Tiantian Yuan for helpful discussions. 

This work is based on observations carried out under project number D06-18 with the IRAM NOEMA Interferometer [30\,m telescope]. IRAM is supported by INSU/CNRS (France), MPG (Germany) and IGN (Spain).
We would like to thank K. Schuster, the director of IRAM, for the generous attribution of Director's Discretionary Time, and the staff and pool observers at the 30-m telescope, in particular C. Kramer and Wonju Kim, for the rapid scheduling and execution of our CO observations. 

The data presented herein were obtained at the W. M. Keck Observatory, which is operated as a scientific partnership among the California Institute of Technology, the University of California and the National Aeronautics and Space Administration. The Observatory was made possible by the generous financial support of the W. M. Keck Foundation.
Australian community access to the Keck Observatory was supported through the Australian Government's National Collaborative Research Infrastructure Strategy, via the Department of Education and Training, and an Australian Government astronomy research infrastructure grant, via the Department of Industry and Science.
We would like to thank the Keck support staff, in particular J. Lyke, for their help in obtaining our OSIRIS observations.

The authors recognize and acknowledge the very significant cultural role and reverence that the summit of Maunakea has always had within the indigenous Hawaiian community. We are most fortunate to have the opportunity to conduct observations from this mountain.

This study uses data provided by the Calar Alto Legacy Integral Field Area (CALIFA) survey (\url{http://califa.caha.es/}), based on observations collected at the Centro Astron\'omico Hispano Alem\'an (CAHA) at Calar Alto, operated jointly by the Max-Planck-Institut f\"or Astronomie and the Instituto de Astrof\'isica de Andaluc\'ia (CSIC). We also thank Sebastian S\'anchez for assistance with accessing CALIFA data products used in this work.

This study made use of data made available by the NASA/IPAC Extragalactic Database (NED), which is operated by the Jet Propulsion Laboratory, California Institute of Technology, under contract with the National Aeronautics and Space Administration.

This research made use of QFitsView\footnote{\url{https://www.mpe.mpg.de/~ott/QFitsView/}}, a software package for reducing astronomical data written by Thomas Ott, 
Scipy\footnote{\url{http://www.scipy.org/}}~\citep{Scipy2001}, and Astropy,\footnote{\url{http://www.astropy.org}} a community-developed core Python package for Astronomy~\citep{Astropy2013,Astropy2018}.


\bibliographystyle{mnras}
\bibliography{bibliography}



\appendix
\section{The morphology of the radio source}
\label{appendix: The morphology of the radio source}

\begin{figure*}
	\centering
	\subcaptionbox{\label{fig: VLBI: 1.665 GHz}}{\includegraphics*[width=0.28\linewidth]{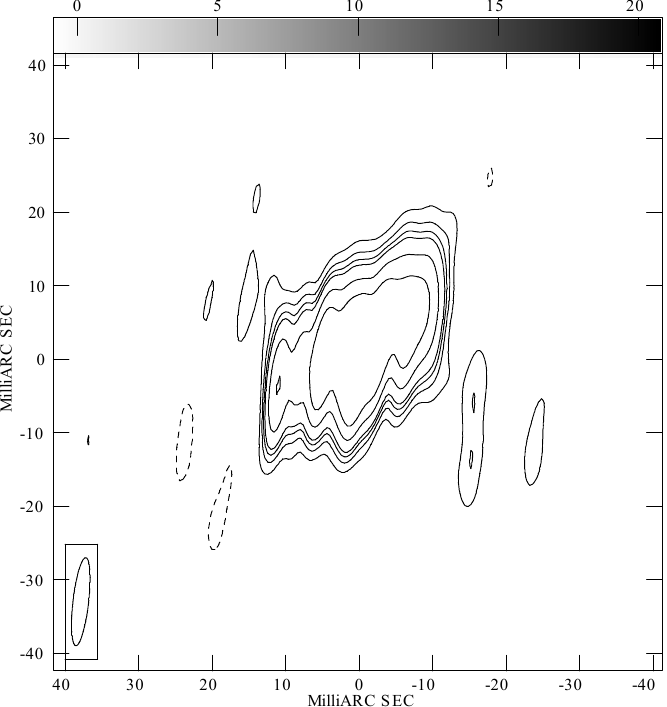}}
	\hfill
	\subcaptionbox{\label{fig: VLBI: 4.993 GHz}}{\includegraphics*[width=0.28\linewidth]{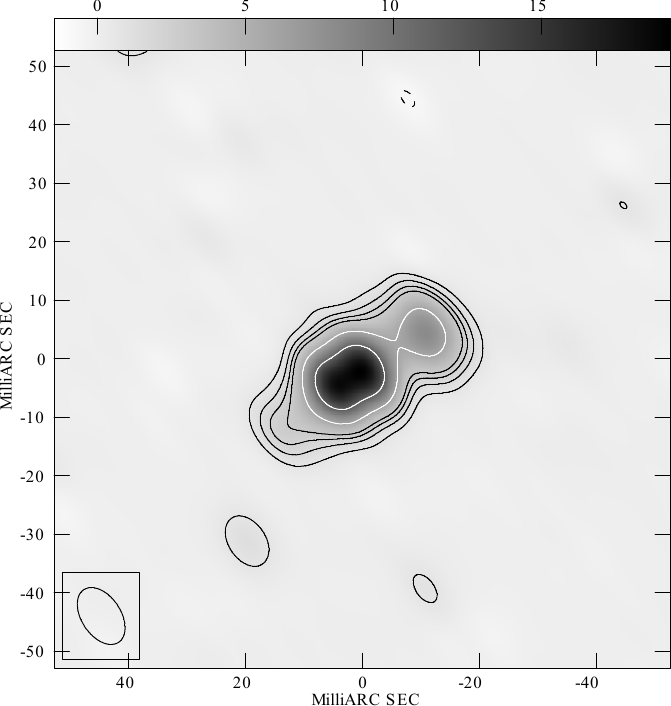}}
	\hfill
	\subcaptionbox{\label{fig: VLBI: 8.4 GHz}}{\includegraphics*[width=0.28\linewidth]{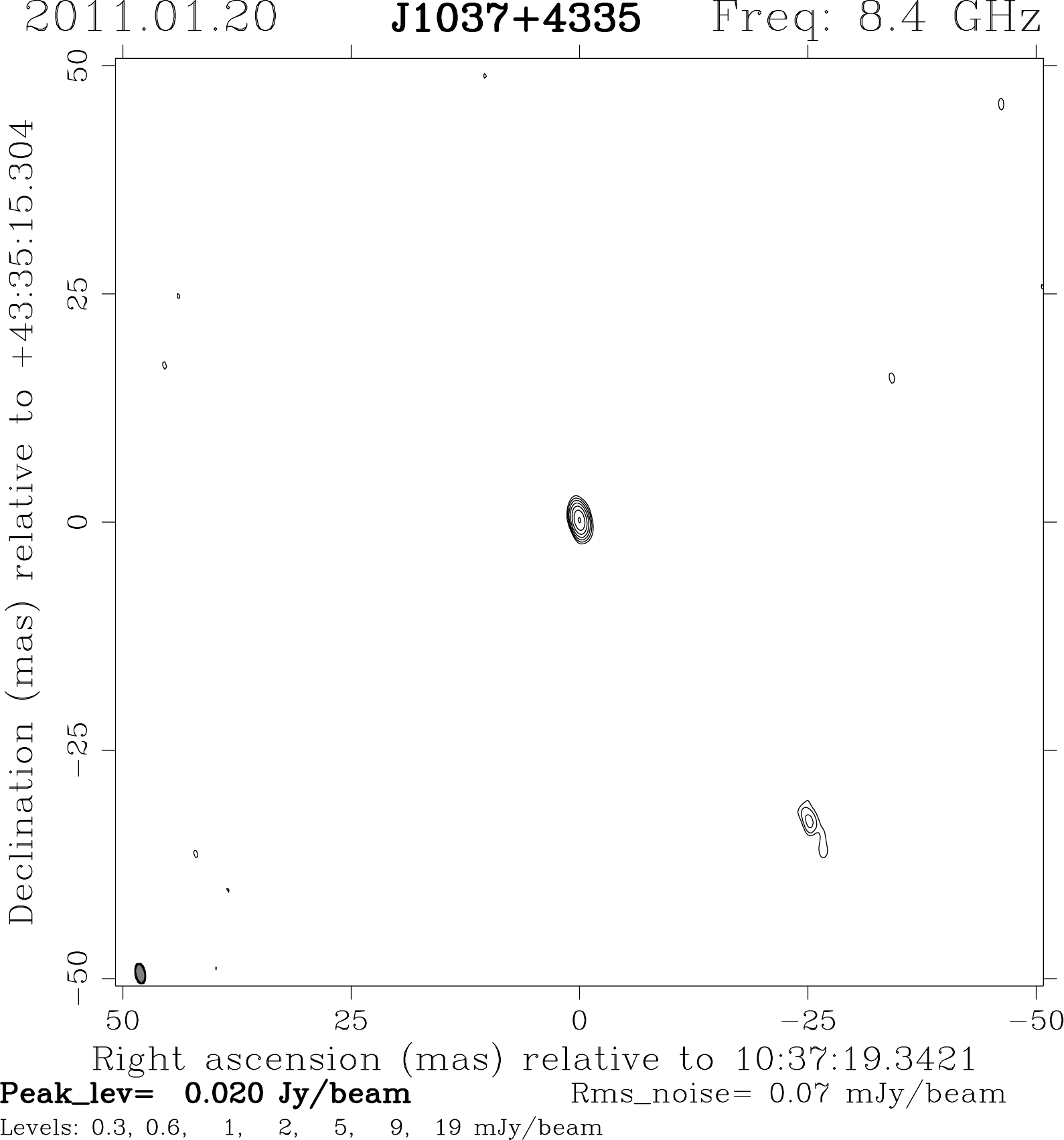}}
	\caption{(a) and (b) show VLBI images of UGC\,05771 at 1.665\,GHz and 4.993\,GHz respectively (\citet{deVries2009}, A\&A, 498, 641, 2009, reproduced with permission \textcopyright{} ESO). (c) shows the VLBI image at 8.4\,GHz, obtained from the astrogeo database\protect\footnotemark. The root-mean-square (rms) sensitivities for these observations are $0.10$, $0.23$ and $0.09\rm\,mJy\,beam^{-1}$ respectively.}
\end{figure*}

The radio source associated with UGC\,05771 has a strikingly different morphology at 1.665\,GHz (Fig.~\ref{fig: VLBI: 1.665 GHz}), 4.993\,GHz (Fig.~\ref{fig: VLBI: 4.993 GHz}) and at 8.4\,GHz (Fig.~\ref{fig: VLBI: 8.4 GHz}). 
A South-West (SW) component at a projected distance of $18.3\,\rm\,pc$ is visible at 8.4\,GHz; whilst there is faint contour to the SW at 4.993~GHz, it is absent at 1.665\,GHz. 
In this section we address the apparent discrepancies between the three images.

The SW contour is offset by $\approx 16\,\rm mas$ ($7.1\,\rm\,pc$) between the 4.993\,GHz and 8.4\,GHz images.
The observations are separated by 6.6\,yr, corresponding to an apparent speed of $\approx 3.5c$, which would only be plausible if the source were strongly beamed. We rule this out due to the source's low flux variability~\citep{Snellen2004}; therefore the SW contour in the 4.993\,GHz image is probably an artefact.

Using the 8.4~GHz radio source count of \citet{Fomalont2002} we estimate the likelihood that the SW component is a background source to be vanishingly small, and therefore is most probably associated with UGC\,05771.
It is instead likely to be a site where jet plasma has interacted with a particularly dense cloud in the ISM, re-accelerating electrons and creating a temporary hot spot.

The SW component may be absent from the 1.665\,GHz and 4.993\,GHz images due to beam smearing. If the integrated flux of a point source is lower than the rms noise per beam, the component will be indistinguishable from noise.
The integrated flux of the SW component at 8.4\,GHz is $S_{\rm int}(8.4\rm\,Ghz) = 2.13\,\rm mJy$~\citep{Cheng2018}. Assuming the spectral index of the source $\alpha = 0.62$~\citep{Snellen2004}, then the integrated fluxes $S_{\rm int}(5.0\rm\,Ghz) = 2.94\,\rm mJy$ and $S_{\rm int}(1.7\rm\,Ghz) = 5.74\,\rm mJy$.
The rms noise of the observations is $0.23\,\rm mJy \, beam^{-1}$ and $0.10\,\rm mJy \, beam^{-1}$ at 4.993\,GHz and 1.665\,GHz respectively, meaning that the SW component would be detected with a high S/N in either case; hence beam smearing cannot explain why the component is not visible. 

We conclude that either (i) the SW component is strongly absorbed at these frequencies, either by synchrotron self-absorption or FFA, or (ii), there is structure at these lower frequencies that has been resolved out, which is consistent with the low flux completeness of these observations ($\leq 30\rm \,per\,cent$, see Fig.~\ref{fig: radio spectrum}).
For example, the jet structure visible at 1.7~GHz and 5~GHz is absent at 8.4\,GHz. Using the flux along the jet axis from Fig.~\ref{fig: VLBI: 4.993 GHz} and assuming $\alpha = 0.62$, the emission at 8.4\,GHz should be $\approx 5\,\rm mJy$, well above the rms noise ($0.07\,\rm mJy$), indicating that the jets have been resolved out in these observations.
It is therefore plausible that the SW component has lower frequency counterparts that are not visible in these observations.

\footnotetext{The astrogeo database is maintained by Leonid Petrov, and is available at \url{http://astrogeo.org/}.}


\bsp	
\label{lastpage}
\end{document}